\documentclass[aps,prd,twocolumn,preprintnumbers,nofootinbib]{revtex4-1}
\usepackage{graphicx}
\usepackage{enumitem}
\usepackage[FIGTOPCAP]{subfigure}
\usepackage{dcolumn}
\usepackage{bm}
\usepackage{epstopdf}
\usepackage{amsmath,amssymb}
\usepackage{bigints}
\usepackage{mathtools}
\usepackage{array}
\usepackage{bbm}
\usepackage{dsfont}
\usepackage{quantikz}
\usepackage[normalem]{ulem} 
\usepackage[breaklinks=true,colorlinks,citecolor=blue,linkcolor=blue,urlcolor=blue]{hyperref}
\usepackage{xcolor}
\usepackage{hhline}
\usepackage{multirow}

\newcommand{\mel}[3]{{\langle {#1}|{#2}|{#3}\rangle}}
\newcommand{\ketbra}[2]{{\ket {#1}\!\!\bra{#2}}}
\newcommand{\ketregbra}[3]{{\ket {#1}_{#2}\!\bra{#3}}}

\begin{document}

\title{Improving Ground State Accuracy of Variational Quantum Eigensolvers with Soft-coded Orthogonal Subspace Representations}

\author{Giuseppe Clemente}\email{giuseppe.clemente@unipi.it}
\author{Marco Intini}\email{marco.intini@phd.unipi.it}
\affiliation{Dipartimento di Fisica dell'Universit\`a di Pisa and INFN - Sezione di Pisa,\\ 
Largo Pontecorvo 3, I-56127 Pisa, Italy.}

\date{\today}

\begin{abstract}
	We propose a new approach to improve the accuracy of ground state estimates
	in Variational Quantum Eigensolver (VQE) algorithms by
	employing subspace representations with soft-coded orthogonality constraints.
	As in other subspace-based VQE methods, such as the
	Subspace-Search VQE (SSVQE) and Multistate Contracted VQE (MCVQE),
	once the parameters are optimized to maximize the subspace overlap with the low-energy sector of the Hamiltonian,
	one diagonalizes the Hamiltonian restricted to the subspace.
	Unlike these methods, where \emph{hard-coded} orthogonality constraints are enforced
	at the circuit level among the states spanning the subspace,
	we consider a subspace representation where orthogonality is \emph{soft-coded} via
	penalty terms in the cost function.
	We show that this representation allows for shallower quantum circuits while maintaining high
	fidelity when compared to single-state (standard VQE) and multi-state (SSVQE or MCVQE) representations,
	on two benchmark cases: a $3\times 3$ transverse-field Ising model
	and random realizations of the Edwards--Anderson spin-glass model on a $4\times 4$ lattice.
\end{abstract}

\maketitle

\section{Introduction}\label{sec:intro}
The search for the ground state of closed quantum systems is one of the most relevant and promising applications of quantum computing, especially in the form of Variational Quantum Eigensolvers (VQE)~\cite{Peruzzo:2013bzg,TILLY20221},
which makes it accessible to the current Noisy Intermediate-Scale Quantum (NISQ) era~\cite{Preskill2018quantumcomputingin}
where other approaches, such as Quantum Phase Estimation~\cite{10.1098/rspa.1998.0164,PhysRevLett.83.5162},
are not feasible due to hardware limitations and sensitivity to noise.

In the standard VQE algorithm, the solution to the ground state search
is realized by means of a single parameterized circuit, the so-called \emph{ansatz},
whose parameters are classically optimized to minimize the expectation value of the Hamiltonian.
On the other hand, there is a wide class of algorithms based on the general idea of embedding the
problem into a larger search space, where excited states are involved. In particular, we consider algorithms
extending the variational search with a single parameterized state vector
into a search involving multiple parameterized state vectors.
This is the case of the Subspace-Search VQE (SSVQE)~\cite{PhysRevResearch.1.033062} and
the Multistate Contracted VQE (MCVQE)~\cite{PhysRevLett.122.230401},
which both share a minimization of the average energy over a subspace
built as the span of multiple parameterized states ($K$-frame) whose orthogonality
is \emph{hard-coded} into the circuit structure itself.
The ideal minimum of the subspace energy is reached when the parameterized subspace
overlaps as best as possible with the low-energy sector of the Hamiltonian.
Once convergence of the loss function is reached,
from the optimal subspace solution one can then extract an approximation to
the ground state (and possibly also higher-energy states)
via exact diagonalization of the Hamiltonian truncated to the optimal subspace.
The approach proposed in this work shares the same overall idea of subspace search,
but differs in the way orthogonality between the states forming the $K$-frame is imposed,
which involves \emph{soft-coded} constraints via penalty terms in the cost function,
as described in Section~\ref{subsubsec:subspace-soft}.

It is also worth mentioning other similar approaches of subspace search
where non-orthogonal frames~\cite{Huggins_2020} are considered and where a
generalized eigenvalue problem is classically solved at each step of the optimization run.
Many of the approaches involving excited states rely on exploiting known properties
of the Hamiltonian (e.g., symmetries) in the construction of the ansatz for the single
parameterized states spanning the subspace, sometimes in the form of Hamiltonian-inspired ans\"atze:
this is the case, for example, of Krylov-based approaches~\cite{Yu:2025czp,Piccinelli:2025gmh}
or approaches related to ADAPT-VQE~\cite{grimsley2019adaptive}.
However, in this work, we focus the analysis on cases that satisfy
the following two fundamental assumptions:
\begin{enumerate}[label=\textbf{A.\arabic*}]
	\item \label{it:agnostic} the ansatz is \emph{agnostic} with respect to the Hamiltonian symmetries or other known properties,
	      such that no prior information is injected into the circuit structure or as penalty terms in the cost function;
	\item \label{it:parallel} the structure of the ansatz is kept fixed during the optimization,
	      without increasing the subspace dimension or changing the circuit besides the values of its parameters.
\end{enumerate}
Assumption~\ref{it:agnostic} is particularly relevant for low-symmetry systems
such as general realizations of spin glasses,
which are usually mentioned as potential candidates for quantum advantage,
especially in the context of quantum annealing approaches~\cite{PhysRevX.5.031026,tanaka2017quantum,chowdhury2025pushing}.
These types of systems motivate the analysis of the approach without using any information about possible symmetries.
Therefore, in Section~\ref{subsec:numres-TFI2D}, 
  we consider as a benchmark the 2D Transverse-Field Ising (TFI) model, 
but without exploiting information on the system.
Another motivation for considering non-symmetric ans\"atze is that enforcing problem-specific symmetries into the ansatz makes it
typically deeper and possibly not feasible for NISQ era applications.
Assumption~\ref{it:parallel}, on the other hand, makes the analysis
and comparison between different subspace representations fairer.
This does not prevent future extensions and integrations of the approach proposed here
with other approaches based on sequential improvements of the ansatz
(relaxing assumption~\ref{it:parallel})
or with the exploitation of system-specific properties
(relaxing assumption~\ref{it:agnostic}),
but these possible variations would not enrich the discussion on the advantages of the soft-coded
representation and are left for future investigations.

The structure of this work is the following:
in Section~\ref{sec:subspace}
we discuss an overview of the standard representation
to approximate ground states for VQE and introduce subspace representations
in terms of frames of states with either hard-coded or soft-coded orthogonality.
In Section~\ref{sec:numres}
we present numerical results on two benchmark models in (2+1) dimensions: a transverse-field Ising model~\ref{subsec:numres-TFI2D}
and an Edwards--Anderson (EA) model (spin glasses)~\ref{subsec:numres-spin-glass}.
Finally, in Section~\ref{sec:concl}
we summarize the results and discuss future perspectives.

\section{Subspace Representations}\label{sec:subspace}
The main goal of the standard formulation of VQE~\cite{Peruzzo:2013bzg}
consists in finding the best approximation to the ground state of a given Hamiltonian $H$.
The search space is represented by single pure states,
which are realized as a parameterized unitary quantum circuit $U(\bm \theta)$
applied to a fixed initial state $\ket{\varphi_0}$,
i.e., $\ket{\psi(\bm \theta)} = U(\bm \theta) \ket{\varphi_0}$.
Then, the parameters ${\bm \theta}$ are optimized classically to minimize
the expectation value of the Hamiltonian $\langle \psi(\bm \theta) | H | \psi(\bm \theta) \rangle$,
whose value is computed on a quantum processor.
Subspace-based extensions of the VQE algorithm mentioned in the previous Section
involve instead the use of multiple parameterized states $\{\ket{\psi_i}\}_{i=0}^{K-1}$,
which generalize the problem of finding the ground state
to the problem of finding the lowest-energy subspace of fixed dimension $K$.
In these cases, the search space is represented by a $K$-dimensional
parameterized subspace $\mathcal V$ of the Hilbert space $\mathcal H$.
In these cases, the cost to be minimized contains the average energy of the subspace but,
while in MCVQE and SSVQE the orthogonality of the set of search states
is imposed as a hard constraint by circuit design,
we propose a soft-coded orthogonality implemented through penalty terms in the cost function,
inspired by Variational Quantum Deflation (VQD)~\cite{Higgott2019variationalquantum}.
Only after convergence is reached, then one formulates
the Hamiltonian eigenvalue problem restricted to this subspace,
diagonalizing it classically, as in MCVQE,
or through an additional VQE search into the subspace,
as in one of the variants of SSVQE\footnote{In the case of SSVQE, a
	few variants of the algorithm are proposed in the original paper.
	However, for the sake of the discussion here,
	we refer to the case where the cost function is a uniform sum of the expectation values of
	the Hamiltonian on all the $K$ states which form a basis for the subspace, 
	while the subspace diagonalization is performed by parameterizing the 
	preparation layer as spanning all linear combinations of these basis states.}.
In the following Sections, we describe in detail the representations of
subspace ans\"atze where the orthogonality between the $K$ states
forming a basis of the parameterized subspace
is either hard-coded into the ansatz (Section~\ref{subsubsec:subspace-hard})
or soft-coded into the cost function (Section~\ref{subsubsec:subspace-soft}).
A discussion of the final diagonalization step is mentioned in Section~\ref{subsec:subspace-meas},
while the expected difference in expressibility between the two representations is discussed
in Section~\ref{subsec:subspace-express}.

\subsection{Subspace Ans\"atze and Optimization}\label{subsec:subspace-ansatze}
With the term \emph{subspace ansatz}, we refer to any parameterized representation
which generates a possibly general $K$-dimensional subspace $\mathcal V$ of the Hilbert space $\mathcal H$.
Furthermore, as mentioned in the assumption~\ref{it:parallel} of the previous Section,
we investigate specifically those representations where the subspace $\mathcal V$, 
generated by the span of a set of $K$ parameterized states,
is optimized as a whole and where \mbox{$\dim \mathcal V = K$} is a fixed hyperparameter,
thus excluding from the present analysis approaches where the subspace is built sequentially
with increasing dimension of the search space,
a scenario which might be investigated as future outlook.

There are different ways to construct such a representation, depending on whether and how
constraints are imposed on the $K$ states that form a basis for $\mathcal V$.
In the agnostic setting considered here, we are mainly concerned with the orthogonality constraints,
which guarantee linear independence of the $K$ states, but other constraints
(for example restricting the search into a specific symmetry sector of the Hamiltonian)
can be imposed as well, as they are certainly compatible with the following discussion.
However, in this work, we do not consider non-orthogonal frame representations
because, without imposing any other constraint on the states forming the subspace,
we observed an effective reduction of the subspace dimension caused by overlaps
between different states getting very close to 1 at convergence,
making them virtually not linearly independent anymore\footnote{Even
	with exact linear independence, highly overlapping states are harder to distinguish
	than orthogonal states. This reflects into a higher number of shots required
	to estimate the solution to the generalized eigenvalue problem at fixed precision
	(i.e., poorly conditioned matrices).}.
Indeed, in the NOVQE approach~\cite{Huggins_2020},
non-orthogonal frames are built as sequential improvements of the ansatz,
which therefore violates the assumption~\ref{it:parallel} stated above.
Moreover, this would involve a diagonalization step at each iteration of the optimization
but, as mentioned in Section~\ref{subsec:subspace-meas},
estimating the off-diagonal matrix elements of the Hamiltonian
between non-orthogonal states is computationally demanding in terms of
circuit executions. For this reason, in our implementation we only perform
the diagonalization step at the end of the optimization run,
where the subspace is less contaminated by high-energy components
(see discussion to Figure~\ref{fig:spectrograms-ising}
in Section~\ref{subsec:numres-TFI2D}).

\subsubsection{Hard-coded Orthogonal Frames}\label{subsubsec:subspace-hard}
In this subspace representation,
used in the SSVQE~\cite{PhysRevResearch.1.033062}
and MCVQE~\cite{PhysRevLett.122.230401} approaches,
the orthogonality between the $K$ basis states is enforced directly at the circuit level.
Indeed, the first step consists in choosing $K$ exactly orthonormal initial states
$\{|\varphi_p\rangle\}_{p=0}^{K-1}$, for example, the first $K$ states
of the computational basis in binary representation,
or the states from a Configuration Interaction basis of single-particle orbitals for Quantum Chemistry applications.
To each of these states, one builds a hard-coded orthogonal $K$-frame $\{\ket{\psi_p(\bm \theta)}\}_{p=0}^{K-1}$
by applying the same parameterized circuit $U(\bm \theta)$,
so that $\ket{\psi_p(\bm \theta)} = U(\bm \theta) |\varphi_p\rangle\; \forall p=0,\dots,K-1$.
The subspace is then defined as the image of the orthogonal projector onto the subspace
\begin{align}
	\mathbb{P}_{\mathcal V_{\bm \theta}} := U(\bm \theta)\Big\lbrack\sum_{p=0}^{K-1} \ketbra{\varphi_p}{\varphi_p}\Big\rbrack U^\dag(\bm \theta).
\end{align}
The cost function, minimized during the optimization,
is proportional to the average energy of the parameterized subspace:
\begin{align}\label{eq:cost-hard}
	C_{K}(\bm \theta) :=  {\rm Tr}\lbrack H \mathbb{P}_{\mathcal V_{\bm \theta}}\rbrack = \sum_{p=0}^{K-1} \mel{\psi_p(\bm \theta)}{H}{\psi_p(\bm \theta)},
\end{align}
where the single expectation values in the sum can be computed with different circuits
or by using a single circuit with additional ancilla qubits~\footnote{For
	a relatively small number of states $K \ll \dim \mathcal H$,
	there is no significant drawback in measuring all the $K$ expectation values separately,
	while building a single circuit with ancilla qubits involves additional controlled operation
	that might be challenging, especially if applied at each iteration of the optimization.
	Moreover, considering a fixed precision in the estimate of the subspace-averaged energy,
	this latter version shows no advantage over the former, requiring the same total number of shots.}.

Once convergence is reached, the best approximation to the ground state
is determined by solving the eigenvalue problem for the Hamiltonian restricted to the subspace.
For example, in a variant of the SSVQE algorithm (described in Section IIA of Ref.~\cite{PhysRevResearch.1.033062}),
this is achieved by freezing the final optimal parameters $\bm \theta^*$
and parameterizing the preparation layer as spanning all linear combinations of the $K$ basis states
through a parameterized preparation $\ket{\varphi(\bm \alpha)}=W(\bm \alpha)\ket{0} = \sum_{p=0}^{K-1} c_p(\bm \alpha) \ket{\varphi_p}$,
realized with a circuit $W(\bm \alpha)$ which possibly guarantees
a full coverage of the $K$-dimensional subspace:
${\cup_{\bm \alpha} \{\ket{\varphi(\bm \alpha)}\}} = {\rm Span} \{\ket{\varphi_p}\} = U^\dag(\bm \theta^*) \cdot \mathcal V_{\bm \theta^*}$.
The solution is found by minimizing the energy of
the single pure state $\ket{\Psi(\bm \alpha)} := U(\bm \theta^*)\ket{\varphi(\bm \alpha)}=U(\bm \theta^*)W(\bm \alpha)\ket{0}$
with respect to the parameters $\bm \alpha$.
Another possibility consists in directly estimating the matrix elements
\mbox{$A_{pq}:=\mel{\psi_p(\bm \theta^*)}{H}{\psi_q(\bm \theta^*)}$}
of the Hamiltonian truncated to the optimal subspace $\mathcal V_{\bm \theta^*}$,
which is the approach considered in MCVQE~\cite{PhysRevLett.122.230401}.
In our numerical tests, we opted for this latter approach,
whose realization is described in detail in Section~\ref{subsec:subspace-meas}.

\subsubsection{Soft-coded Orthogonal Frames}\label{subsubsec:subspace-soft}
Here we consider a subspace representation
where orthogonality between the $K$ basis states is not strictly enforced at the circuit level,
but is simply penalized via additional terms in the cost function.
Without loss of generality, one can start from a single
fixed initial state $\ket{0}$ but, unlike the hard-coded version,
states of the $K$-frame are built by applying (possibly distinct)
independent parameterized circuits $\{U_p(\bm \theta_{(p)})\}_{p=0}^{K-1}$,
where the parameters $\bm \theta_{(p)}$ and $\bm \theta_{(q)}$ with $p\neq q$ are unrelated.
The total number of parameters can be written for simplicity as
a ordered tuple $\bm \theta := (\bm \theta_{(0)}, \bm \theta_{(1)}, \ldots, \bm \theta_{(K-1)})$.
The states of the $K$-frame are therefore of the form $|\psi_p(\bm \theta_{(p)})\rangle = U_p(\bm \theta_{(p)})\ket{0}$
and their span defines the subspace $\mathcal V_{\bm \theta}:={\rm Span}\{|\psi_p(\bm \theta_{(p)})\rangle\}_{p=0}^{K-1}$
but, unlike the hard-coded orthogonal case,
the sum of the single state projectors $\ketbra{\psi_p}{\psi_p}$
is not guaranteed to form an orthogonal projector into the subspace.
The cost function to be minimized in the soft-coded orthogonal case reads
\begin{align}\label{eq:cost-soft}
	C_K(\bm \theta) & = \sum_{p=0}^{K-1} \mel{\psi_p(\bm \theta_{(p)})}{H}{\psi_p(\bm \theta_{(p)})} \nonumber \\
	                & + \beta \sum_{p < q} \left|\braket{\psi_q(\bm \theta_{(q)})}{\psi_p(\bm \theta_{(p)})}\right|^2,
\end{align}
and involves both a term proportional to
the average energy of the single states in the frame
(which, can slightly deviate from the exact subspace average
${\rm Tr}\lbrack H \mathbb P_{\mathcal V_{\bm \theta}}\rbrack$ in
case of non exactly orthogonal frames),
and a deflation term,
controlled by a hyperparameter $\beta > 0$
penalizing pairwise overlaps between different states of the frame,
similarly to the one used in Ref.~\cite{Higgott2019variationalquantum}.
Therefore, orthogonality is here enforced only \emph{softly}, i.e.,
as a penalizing term in the cost function,
but it is not strictly enforced during the optimization.
It is worth mentioning that using too small or too large values of $\beta$ might lead to sub-optimal results:
small values would not avoid overlaps between different states in the frame,
possibly giving rise to linear dependence and ill-conditioned eigenvalue problems.
On the other hand, very large values of $\beta$ would mostly 
focus the optimization on heavily minimizing overlaps rather than the energy,
leading to worse results
(see Appendix C of Ref.~\cite{PhysRevD.106.114511}
for a discussion on this aspect in a VQD context).

As in the previous hard-coded case,
once convergence of the cost is reached,
the best approximation to the ground state is determined by solving
an eigenvalue problem for the Hamiltonian restricted
to the subspace.
The main difference with respect to the hard-coded case is that,
besides off-diagonal matrix elements of the Hamiltonian
\mbox{$H_{pq} = \mel{\psi_p(\bm \theta^*_{(p)})}{H}{\psi_q(\bm \theta^*_{(q)})}$},
it is generally necessary to estimate also the overlap matrix elements
\mbox{$S_{pq} = \langle{\psi_p(\bm \theta_{(p)}^*)}|{\psi_q(\bm \theta_{(q)}^*)}\rangle$},
making this a generalized eigenvalue problem: $H\bm c = \lambda S \bm c$
(see next Section for details).
In general, the circuits representing $U_p$ might be different
(for example, with different depths or structure),
but, as discussed in Section~\ref{subsec:subspace-meas},
it is easier to setup a protocol that estimates the off-diagonal terms in the Hamiltonian
truncated to the subspace if we consider them to the equal to the same parameterized circuit $U$.
The crucial difference between the subspace representation discussed here and the
hard-coded orthogonal representation discussed in the previous Section~\ref{subsubsec:subspace-hard}
is that parameters steering each frame state are independent in the soft-coded orthogonal case.

\subsection{Estimate of Hamiltonian Off-Diagonal Terms}\label{subsec:subspace-meas}
As mentioned in the previous Sections on hard-coded and soft-coded orthogonal frames,
once convergence is reached (the dependence on $\bm \theta^*$ is understood in the following),
the optimal approximation to the ground state is obtained by diagonalizing
the Hamiltonian restricted to the subspace:
\mbox{$H_{\rm trc} := H\big\rvert_{\mathcal V_{\bm \theta^*}}$}.
This requires estimating the off-diagonal matrix elements
\mbox{$(H_{\rm  trc})_{pq} := \mel{\psi_p}{H}{\psi_q}$ for $p < q$},
with the addition of the overlap matrix elements
\mbox{$S_{pq} := \braket{\psi_p}{\psi_q}$} in the case of soft-coded orthogonal frames.
These matrix elements are complex numbers in general,
so, to estimate them one needs to measure both their real and imaginary parts
with separate circuits\footnote{It is actually possible to measure both real and imaginary
	parts with a single circuit by using an additional qubit but,
	at fixed level of statistical precision on the estimates,
	this procedure does not provide any advantage with respect to the protocol with two separate circuits.}.
For hard-coded orthogonal subspace ansätze,
the Hamiltonian off-diagonal elements is achieved by preparing superposition states of the form
\begin{equation}
	\begin{aligned}
		\ket{\Upsilon_b^{(pq)}} & :=\tfrac{1}{\sqrt{2}}(|\psi_p\rangle + i^b|\psi_q\rangle)                                          \\
		                        & =U(\bm \theta^*) \tfrac{1}{\sqrt{2}}(|\varphi_p\rangle + i^b|\varphi_q\rangle), \quad b\in \{0,1\}
	\end{aligned}
\end{equation}
directly on the system register.
By measuring the expectation value of
the Hamiltonian on these states for any pair $p<q$,
one can build the matrix associated with the truncated Hamiltonian,
as described in Refs~\cite{PhysRevResearch.1.033062,PhysRevLett.122.230401}.
Since the overlap matrix is the identity ($S = \mathds{1}$),
this final diagonalization step is a standard dense eigenvalue problem
which can be solved on a classical computer.

Regarding soft-coded orthogonal subspace ansätze,
we already mentioned that an estimate of the matrix elements of the overlap matrix
$S_{pq} = \braket{\psi_p}{\psi_q}$ is required for correctness.
Both $H_{\rm trc}$ and $S$ are measured using an ancilla qubit
and a Hadamard-based test protocol as described next.

\begin{figure} 
	\centering
	\includegraphics[width=1.0\linewidth]{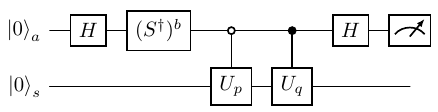}
	\caption{Generalized Hadamard test circuits for the estimate of real and imaginary parts
		of off-diagonal matrix elements between pure states $\ket{\psi_p}:=U_p \ket{0}$ and $\ket{\psi_q}:=U_q \ket{0}$.
		At the end of the circuit, the system register is either traced out (ignored) for the estimate of the overlap matrix elements $S_{pq}$,
		or measured to estimate of the Hamiltonian matrix elements $(H_{\rm trc})_{pq}$ (see text).
		The ancilla and system registers are labeled by `$a$' and `$s$' respectively.}
	\label{fig:hadamard_test}
\end{figure}

Although the parameters $\bm \theta_{(p)}$ are independent for distinct $p$,
for simplicity, we choose the circuit structure to be the same for all states,
i.e., $U_p:=U(\bm \theta_{(p)})$ for a given common parametric circuit $U(\cdot)$.
To prepare the required superposition states, we can use a generalized Hadamard test protocol,
depicted in Figure~\ref{fig:hadamard_test} which can be used for the estimation
of both the overlap matrix elements $S_{pq}$ and the Hamiltonian matrix elements $(H_{\rm trc})_{pq}$.
The open control ($\circ$) applies $U_p$ when the ancilla is in state $\ket{0}$,
while the filled control applies $U_q$ when the ancilla is in state $\ket{1}$,
while the conjugate of a phase gate ($S= \text{diag}(1, i)$ in the computational basis)
is either applied ($b=1 \rightarrow S^\dag$) to extract the imaginary part of the desired matrix elements, or not ($b=0\rightarrow \mathds{1}$) if one is interested in its real part.
After the final Hadamard gate on the ancilla, the output state is
\begin{align}
	|\widetilde{\Upsilon}_b^{(pq)}\rangle = \frac{1}{2}\Big[\ket{0}_a (U_p + (-i)^b U_q)\ket{0}_s + \ket{1}_a (U_p - (-i)^b U_q)\ket{0}_s\Big],
\end{align}
so that, the probability of measuring $m\in\{0,1\}$ on the ancilla qubit, reads
\begin{equation}
	\begin{aligned}
		P(m\mid b, p, q) & := {\rm Tr}\Big[\Big(\ketregbra{m}{a}{m} \otimes \mathds{1}_s\Big)\ketbra{\widetilde{\Upsilon}_b^{(pq)}}{\widetilde{\Upsilon}_b^{(pq)}}\Big] \\
		                 & =\tfrac{1}{2} [1 + (-)^m (\delta_{b,0}{\rm Re} S_{pq}+\delta_{b,1}{\rm Im} S_{pq})].
	\end{aligned}
\end{equation}
We can then estimate both the real and imaginary parts of the overlap
matrix elements by using, respectively, the $b=0$ or $b=1$ variants of the circuit
in Figure~\ref{fig:hadamard_test}:
\begin{align}
	{\rm Re}\, S_{pq} & = 2\, P(m=0 \mid b=0,p,q) - 1, \\
	{\rm Im}\, S_{pq} & = 2\, P(m=0 \mid b=1,p,q) - 1.
\end{align}
For later convenience, we can use the exact eigenbasis $\{\ket{\phi_j^{\rm exact}}\}_j$
of the Hamiltonian $H$ to express the vector components of the $\ket{\psi_p}$
state of the frame as $\psi_p^j :=\braket{\phi_j^{\rm exact}}{\psi_p}$,
from which we can formally write the estimate the off-diagonal matrix elements
of the truncated Hamiltonian \mbox{${(H_{\rm trc})}_{pq}=\mel{\psi_p}{H}{\psi_q}=\sum_j E_j \bar{\psi}_p^j \psi_q^j$},
instead we measure the expectation value of $\widetilde{H} = (\sigma_z)_a \otimes (H)_s$
on the output states $|\Upsilon_{b=0,1}^{(pq)}\rangle$, which results in
\begin{equation}
	\begin{aligned}
		 & \langle \widetilde{H} \rangle_{\widetilde{\Upsilon}_b^{(pq)}}  := \sum_{j,m} (-)^m E_j \big|\bra{m}_a\!\bra{\phi_j^{\rm exact}}_s \ket{\widetilde{\Upsilon}_b^{(pq)}}\big|^2 \\
		 & = \tfrac{1}{4} \sum_{j,m} (-)^m E_j |\psi_p^j + (-)^m (-i)^b \psi_q^j|^2                                                                                                     \\
		 & = \delta_{b,0}{\rm Re} H_{pq}+\delta_{b,1}{\rm Im} H_{pq}                                                                                                                    \\
		 & \implies H_{pq}:=\langle \widetilde{H} \rangle_{\widetilde{\Upsilon}_0^{(pq)}}+i\langle \widetilde{H} \rangle_{\widetilde{\Upsilon}_1^{(pq)}}.
	\end{aligned}
\end{equation}
The soft-coded orthogonality involves a resource trade-off:
while hard-coded representations guarantee $S_{pq} = \delta_{pq}$ by design,
the soft-coded representation requires the explicit estimation of the $K(K-1)/2$
off-diagonal complex elements of $S$.
However, as we show in Sec.~\ref{sec:numres},
the soft-coded representation achieves high fidelities with shallow circuits,
whereas the hard-coded representation saturates at lower fidelities,
failing to reach comparable accuracy even with deep circuits.
This trade-off is particularly advantageous for NISQ hardware,
where minimizing circuit depth is more critical (due to decoherence)
than reducing the number of measurements.
Finally, after reconstructing both $H_{\rm trc}$ and $S$,
the $K\times K$ generalized eigenvalue problem
$H_{\rm trc} \mathbf{c} = \lambda \, S \, \mathbf{c}$,
can be solved via classical routines:
the lowest eigenvalue $\lambda_0$ is
an estimate of the exact ground state energy $E_0$,
while the corresponding eigenvector $\mathbf{c}^*_0$ provides the coefficients
to express an approximation to the ground state
in the form of a linear combination of the optimal parameterized states $\{\ket{\psi_p}\}$
found at convergence:
\begin{align}\label{eq:approx-ground-state}
	\ket{\Psi_0} := \sum_{p=0}^{K-1} c^*_p \, \ket{\psi_p}.
\end{align}
However, the estimated ground energy $\lambda_0 \approx E_0$ might be affected by biases,
while the state $\ket{\Psi_0}$ does not necessarily coincide with the best approximation
within the subspace, as we discuss more thoroughly in the next Section.

\subsection{Subspace and Best Ground State Fidelities}\label{subsec:subspace-fidelities}
Let us consider the eigenvalue problems discussed above, involving
the Hamiltonian $H$ truncated to the subspace \mbox{$\mathcal V\subset \mathcal H$}
for both the hard-coded and soft-coded orthogonal subspace representations.
It is important to stress that the quality of the estimated optimal
approximate ground state $\ket{\Psi_0}$ would depend in general
also on the presence of non-zero off-diagonal blocks
between the subspace $\mathcal V$ and the rest of the Hilbert space $\overline{\mathcal V}:=\mathcal V^\perp$.
To be more precise, let us define the orthogonal projector onto and out of the subspace as
$\mathbb{P}_{\mathcal V}$ and $\overline{\mathbb{P}}_{\mathcal V}:=\mathds{1}-\mathbb{P}_{\mathcal V}$ respectively.
The full Hamiltonian can be decomposed in four blocks as
\begin{align}
	H = \mathbb{P}_{\mathcal V} H \mathbb{P}_{\mathcal V} + \mathbb{P}_{\mathcal V} H \overline{\mathbb{P}}_{\mathcal V}
	+ \overline{\mathbb{P}}_{\mathcal V} H \mathbb{P}_{\mathcal V} + \overline{\mathbb{P}}_{\mathcal V} H \overline{\mathbb{P}}_{\mathcal V},
\end{align}
where we the first term is what we refer to as the \emph{truncated Hamiltonian}
$H_{\rm trc}:= \mathbb{P}_{\mathcal V} H \mathbb{P}_{\mathcal V}$ in the following.
Since off-diagonal blocks, involving mixing terms 
between $\mathcal V$ and $\overline{\mathcal V}$,
are generally non-zero, the eigenstates found from solving the 
eigenproblem with $H_{\rm trc}$ do not coincide with the projection 
of the true eigenstates of $H$ onto $\mathcal V$.
For this reason, whilst common in literature,
we do not find it appropriate to call the restricted operator
as the \emph{effective} Hamiltonian, since this term is usually reserved
to extensions of the truncated Hamiltonian which take into account
the effect of the non-vanishing off-diagonal blocks connecting the subspace ${\cal V}$ with its complement $\overline{\cal V}$.

With these considerations in perspective,
to assess the quality of the approximated ground state when subspaces are involved,
it is useful to introduce two different fidelity metrics:
\begin{itemize}
	\item The \emph{subspace fidelity}:
	      \begin{align}\label{eq:def-subfid}
		      \mathcal F_{\rm sub} := \| \mathbb{P}_{\mathcal V} | \phi_0^{\rm exact} \rangle \|^2,
	      \end{align}
	      measuring the overlap between the exact ground state $\ket{\phi_0^{\rm exact}}$ and the subspace $\mathcal V$.
	      This provides the fidelity for the best possible approximating state produced using only
	      states in the subspace, i.e., $\ket{{\Psi}'_0} := \mathbb{P}_{\mathcal V} | \phi_0^{\rm exact} \rangle / \| \mathbb{P}_{\mathcal V} | \phi_0^{\rm exact} \rangle \|$;
	\item The \emph{truncated} or \emph{$H_{\rm trc}$ fidelity}:
	      \begin{align}\label{eq:def-htrcfid}
		      \mathcal F_{\rm trc} := |\langle \Psi_0 | \phi_0^{\rm exact} \rangle|^2,
	      \end{align}
	      where $\ket{\Psi_0}$ is the solution to the diagonalization
	      of the truncated Hamiltonian $H_{\rm trc}$;
\end{itemize}
By construction, the two fidelities are related by the inequality
\mbox{$\mathcal F_{\rm trc} \leq \mathcal F_{\rm sub}$},
with equality occurring only if the off-diagonal blocks of $H$ between $\mathcal V$
and its orthogonal complement vanish.
Since we cannot access the exact ground state $\ket{\phi_0^{\rm exact}}$ in actual applications,
it is fair to assess only the quality of the output state of the truncated
diagonalization $\ket{\Psi_0}$, which would be
suboptimal to the best state in the subspace $\ket{{\Psi}'_0}$,
for which a knowledge of $\ket{\phi_0^{\rm exact}}$ is required.
The subspace fidelity provides just an ideal upper bound to the achievable
accuracy within the subspace.
For this reason, as a figure of merit for the comparison between different subspace
representations, we only consider the $H_{\rm trc}$ fidelity.
For our numerical tests in Section~\ref{sec:numres},
where we use the exact state as a benchmark but not in the optimization procedure,
we monitor the evolution of both fidelity definitions and their discrepancy
at each iteration.

\subsection{Subspace Expressibility}\label{subsec:subspace-express}
Finally, it is worth discussing the formal differences in expressibility between
the hard-coded and soft-coded orthogonal subspace representations.
In particular, in the hard-coded orthogonal case, the solution has the form
\begin{align}
	\ket{\Psi_{\rm ho}} = U(\bm \theta) \cdot \sum_{p=0}^{K-1} c_p\ket{\varphi_p},
\end{align}
where the coefficients $\{c^p\}$ are found either by classical diagonalization, as described in
the previous Section, or by the optimization of an additional parameterized preparation layer
\mbox{$W(\bm \alpha)\ket{0}:=\sum_{p=0}^{K-1} c_p\ket{\varphi_p}$},
as mentioned in Section~\ref{subsec:subspace-ansatze} for the SSVQE approach.
This means that, from the expressibility point of view,
the hard-coded orthogonal case is equivalent to optimizing
a single state parametric circuit \mbox{$U(\bm \theta) W(\bm \alpha)\ket{0}$},
where the depth of the $W$ layer simply adds to the depth of $U$
without a significant expected advantage in terms of reachable states
and fidelity with respect to the exact ground state.
On the other hand, in the soft-coded orthogonal case
the solution has the form
\begin{align}
	\ket{\Psi_{\rm so}} = \Bigl[\sum_{p=0}^{K-1} c_p U_p(\bm \theta_p)\Bigr]\ket{0} =: V(\{\bm \theta_p\}) \ket{0},
\end{align}
where $V$ is a linear combination of generally non-commuting unitaries.
This representation can therefore generate highly entangled states even with shallow circuits
for $U_p(\bm \theta_p)$, enriching the set of reachable states
and therefore the expected expressibility.
However, while this intuition seems reasonable and is supported by numerical evidence,
as shown in Section~\ref{sec:numres}, a quantitative analysis of the expressibility
of such representations is not straightforward,
since it would require defining proper metrics for the expressibility of single states
within a parametric subspace, which we did not find in literature.
Indeed, most of the literature on expressibility and its relation with barren plateaus focuses
on single state ans\"atze~\cite{ExpressibilitySim2019,Funcke2021dimensional,Arrasmith_2022,PRXQuantum.3.010313,Brozzi:2025omu};
often depending on comparison with Haar-random states or t-designs,
which are not directly applicable to subspace representations, since a non-uniform
coverage of the full Hilbert space is not an issue, as long as the relevant subspace
(e.g., low-energy states) is well represented by
at least some portion of the parameterized subspaces $\cal V_{\bm \theta}$.
As possible metric, we considered a random sampling of
subspaces $\cal V_{\bm \theta}$ for random parameters $\bm \theta$,
which would provide an approximation to the relative coverage of the Hilbert space
$\cup_{\bm \theta} {\cal V}_{\bm \theta}$ but, in practice,
this approach is unfeasible even for systems with very few qubits.
For these reasons, a quantitative analysis of the expressibility
of the considered subspace representations is left for future work.

\section{Numerical Results}\label{sec:numres}
For conciseness, in the rest of this section, we use the shorthands
\emph{hard-ortho} and \emph{soft-ortho} to indicate the subspace ans\"atze with respectively hard-coded orthogonality or soft-coded orthogonality,
as introduced in Sections~\ref{subsubsec:subspace-hard} and~\ref{subsubsec:subspace-soft}.

Both benchmark models considered in this work are represented by Ising-like Hamiltonians
of the form
\begin{align}\label{eq:IsinglikeHam}
	H = -\!\!\!\sum_{(i,j)\in E}\!\!\!\! J_{ij}\, \sigma^x_i \sigma^x_j - h\! \sum_{i\in V} \sigma^z_i,
\end{align}
where $G=(V,E)$ is the undirected graph with vertex set $V$
and neighboring relation $E$ (i.e., the edges),
$J_{ij}$ and $h$ denote variable couplings,
while $\{\sigma^x_i,\sigma^y_i,\sigma^z_i,\}$ are the Pauli operators
acting on the qubit at site $i$.
For our specific runs, we consider the following two models near criticality:
\begin{itemize}
	\item {\bf Transverse-Field Ising (TFI)}: a $3\times 3$ periodic square lattice
	      with homogeneous ferromagnetic coupling $J_{ij}=J=1$ and
	      transverse field $h/J=3.044$,
	      which is close to the critical point of the model~\cite{PhysRevE.66.066110,PhysRevB.102.094101} (at least in the thermodynamic limit);
	\item {\bf Edwards--Anderson (EA)}: a $4\times 4$ periodic square lattice with random couplings $J_{ij}$ independently and identically distributed
	      gaussian variables with zero mean and unit variance, and
	      transverse field $h=2$~\cite{PhysRevLett.72.4141,PhysRevE.110.065305}.
\end{itemize}
We consider here relatively small system sizes,
focusing on a proof-of-concept for the soft-ortho subspace representation.
The TFI model, discussed in Section~\ref{subsec:numres-TFI2D}
provides a first benchmark with relatively well-behaved properties,
while the EA spin-glass model, discussed in Section~\ref{subsec:numres-spin-glass},
provides a more challenging test case 
due to its disordered couplings and lack of exploitable symmetries.
The present study relies on classical emulation,
which limits the accessible system sizes;
larger systems may be explored in follow-up investigations on actual quantum hardware.

\subsection{Numerical Setup}\label{subsec:numres-setup}
\begin{figure} 
	\centering
	\includegraphics[width=1.0\linewidth]{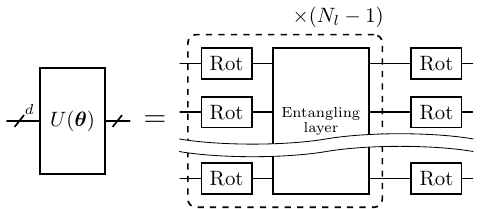}
	\caption{General structure of the parameterized circuit representing $U(\bm \theta)$ in all cases considered. The number of single-qubit independent rotation layers ({\rm Rot}) is $N_l\geq 1$. For $N_l\geq 2$, these are alternated with $N_l-1$ layers of entangling gates (e.g., CNOT, CZ).}
	\label{fig:ansatz}
\end{figure}
Regarding the circuit, for all tests we use a layered structure with
alternating parameterized single-qubit rotations and entangling layers
as generically depicted in Figure~\ref{fig:ansatz},
where we assume the connectivity between qubits
(i.e., availability of 2-qubit gates entangling neighboring sites)
to reflect the 2D lattice model. 
In particular, as single-qubit rotation gates we use ${\rm Rot} = R_z(\beta) R_y(\alpha)$ 
with independently trainable parameters for different qubits. 
Regarding the entangling layer, we use $CZ$ gates 
between all neighboring qubits in the lattices considered.
Thus, for a $q$ qubits circuit, the number of parameter is $N_p = 2\times q\times N_l$.
We choose the same fixed circuit structure $U(\cdot)$ for both the hard-ortho and soft-ortho
frame representations but, while in the former it is applied with common parameters $\bm \theta$
to the first $K$ states of the computational basis (i.e., \mbox{$\ket{\varphi_p}=\ket{{\rm bin}(p)}$}),
for the soft-ortho representation it is always applied to $\ket{0}$,
but with distinct parameters sets $\bm \theta_{(p)}$ for different frame states.

The runs discussed in the following are emulated on classical hardware
using the AerSimulator from the Qiskit library~\cite{qiskit2024} in \emph{statevector} mode,
so that the results are free from both sampling noise and quantum noise.
Considering the case of a finite number of shots or noisy channels would certainly affect
the optimization performance, especially on the final part of the run where variations of the cost function are relatively small.
On the other hand, the stage of the algorithm which we expect to be mostly affected by shot noise
is the estimation of the off-diagonal elements of the Hamiltonian in the subspace (discussed in Section~\ref{subsec:subspace-meas}).
Indeed, at the end of the optimization stage, the subspace would mostly overlap with the low-energy sector of the Hamiltonian,
where energy gaps are small, and accurate estimates of 
off-diagonal elements would be required to correctly and robustly resolve the eigenvalues in the classical diagonalization routine.
In general, an adaptive scheme in the shots allocation, distinguishing between optimization and diagonalization stages,
might be in order to maximize the overall algorithm performance on hardware.
The main goal of this work is to benchmark the advantage in terms of achievable fidelities
of the soft-ortho subspace ansatz over the hard-ortho one and standard VQE,
so we leave the study of these aspects and the exploration of optimal shot schemes
in subspace approaches for future work, focusing here on the ideal shotless case for both stages.

Regarding the optimization loop, we used the NFT optimizer~\cite{Nakanishi:2019rrm}
from Qiskit Algorithms, but different choices would mostly affect
the convergence rates, not the final fidelities achieved.
The parameters for different runs are initialized independently 
and randomly from a uniform distribution in the reduced-domain $[-0.2\pi, 0.2\pi]$.
This choice aligns with common practices in VQE literature,
aiming to avoid potential issues such as barren plateaus,
as well as ensuring an appreciable overlap with the target state,
which is shown to be quite relevant for the convergence of VQE-like algorithms~\cite{zhang2022escaping,PhysRevApplied.22.054005,Arrasmith_2022,PRXQuantum.3.010313,Larocca:2024plh}.
In all cases considered,
we set the maximum number of iterations to $1500$.
Depending on the total number of parameters, 
this might not always be sufficient to reach convergence, as we show in the following.
Furthermore, at each step of the optimization we also compute fidelity metrics
(both subspace fidelity and $H_{\rm trc}$ fidelity, as defined in Section~\ref{subsec:subspace-fidelities}),
which involve using the exact ground state and performing exact diagonalization
of the Hamiltonian restricted to the subspace.
We stress that these intermediate measurements are not used in the optimization process itself,
but are useful for monitoring more fine-grained aspects of the convergence behavior besides
the global loss.
In particular, for the subspace representations,
the matrix elements $\mel{\psi_p}{H}{\psi_q}$
(and the overlap matrix elements $\braket{\psi_p}{\psi_q}$, in the soft-ortho case)
are meant to be measured only as final stage of the run after convergence is reached.

\subsection{Results for Transverse-Field Ising Model}\label{subsec:numres-TFI2D}

In this Section we discuss the specific setup and results regarding the 2D transverse-field Ising model
on a 3×3 lattice with periodic boundary conditions and near criticality,
with $h/J = 3.044$ being
the critical point in the thermodynamic limit~\cite{PhysRevE.66.066110,PhysRevB.102.094101}.
For each \emph{ansatz configuration} (i.e., number of layers $N_l$ and subspace dimension $K$),
we perform 10 optimization runs.
For both hard-ortho and soft-ortho algorithms,
we consider different numbers of layers, as defined in Figure~\ref{fig:ansatz},
and different values of the subspace dimension $K$.
In the soft-ortho case, for all runs we set a penalty factor to $\beta = 10$,
which is sufficient to enforce approximate orthogonality between the states of the frame,
at least in the terminal part of the optimization runs.

In the following Section,
we discuss the relation between the values of the cost
function and the fidelity of ground state estimate during the optimization process
between different subspace approaches and the standard single-state VQE.
In particular, we show how the standard approach is affected by
a non-monotonic behavior of the fidelity during optimization,
providing an argument for its degradation and the advantage of subspace methods.
In Section~\ref{subsubsec:numres-fidelitysummary},
we summarize the final fidelity results after optimization as
a function of the ansatz depth and subspace dimension,
comparing the different approaches.
Finally, in Section~\ref{subsubsec:numres-relative-gain},
we introduce a metric to quantify the relative gain in fidelity
of the subspace approaches with respect to standard VQE,
discussing its behavior as a function of the ansatz depth and subspace dimension.
This, while not being strictly necessary for the TFI model considered here,
would be useful when discussing the EA model in Section~\ref{subsec:numres-spin-glass},
where different disorder realizations lead to different baseline VQE fidelities
and a direct comparison of absolute fidelities would be less meaningful.
It also provides a possibly more general metric to assess the relative advantage between
different approaches independently of the specific model considered.

\subsubsection{Relation Between Cost and Fidelity During Optimization}\label{subsubsec:numres-TFI2D-lossfid}
\begin{figure*}
	\centering
	\includegraphics[width=1\linewidth]{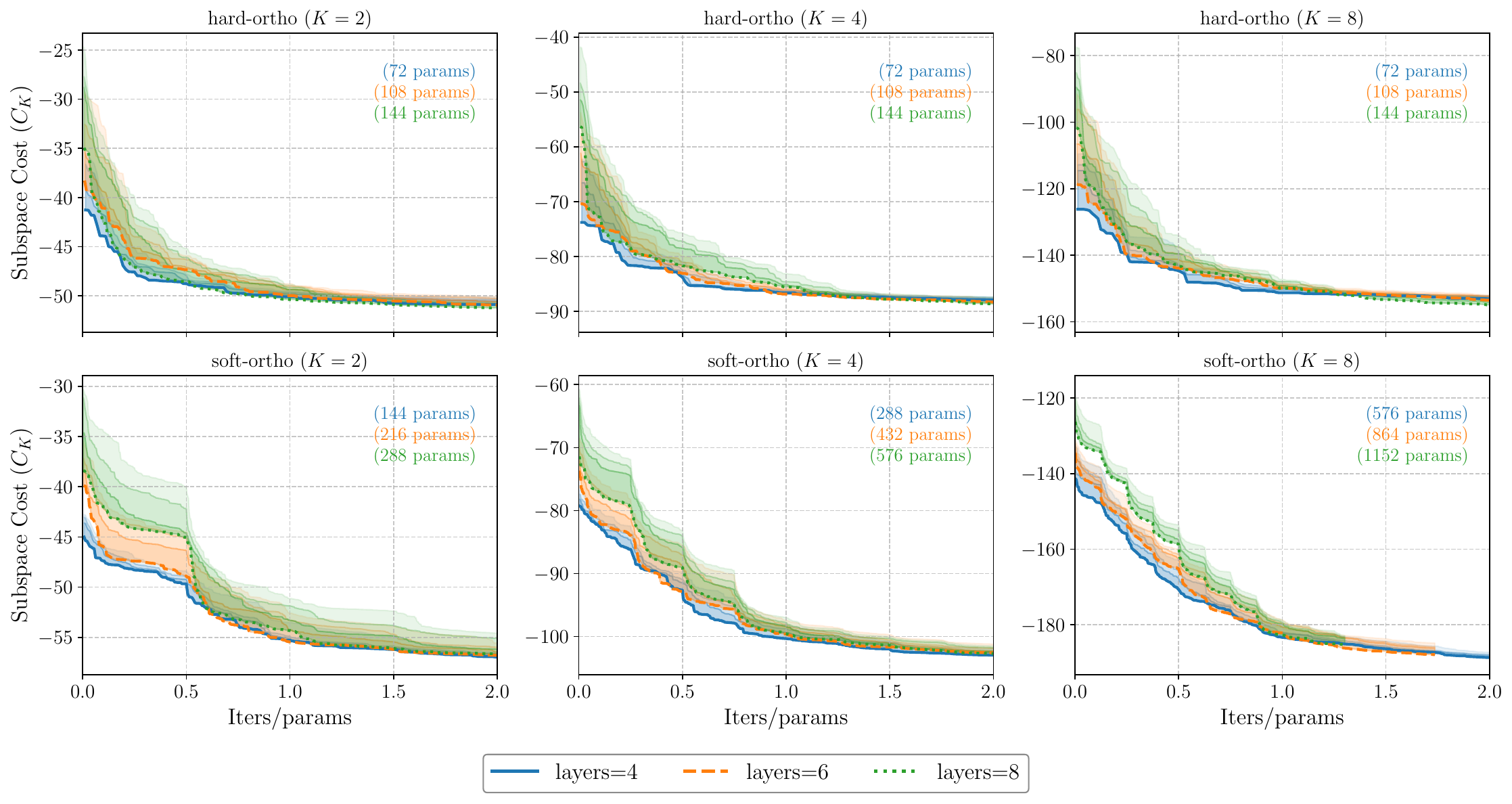}
	\caption{Subspace cost as function of the ratio between optimizer iterations and parameters
		for both hard-ortho (top row) and soft-ortho (bottom row) and for subspace dimension
		$K \in \{2,4,8\}$ for the 2D transverse-field Ising model discussed in the text.
		In each panel, the blue, orange and green curves (solid, dashed
		and dotted, respectively) correspond to ans\"atze for three
		selected depths: $N_l\in \{4,6,8\}$.
		The bottom lines indicate the best costs over 10 repeated
		independent runs with different random initializations,
		while the corresponding shaded bands indicate the spread of the cost values
		from the best run up to the 25th, 50th and 75th percentiles across runs.
		All runs are stopped at 1500 iterations.}
	\label{fig:cost-itovpars-ising}
\end{figure*}
As a first look at the general convergence behavior of the subspace algorithms,
in Figure~\ref{fig:cost-itovpars-ising} we show the subspace cost
as a function of the ratio between the number of optimizer iterations and
the number of trainable parameters for both hard-ortho (top row)
and soft-ortho (bottom row) representations and for different ansatz depths $N_l$.
This choice for the horizontal axis is motivated by the fact that the number of parameters
affects the convergence rate of the optimization but, by rescaling the
number of iterations by the number of parameters, we observe a reasonable
collapse of the curves corresponding to different ansatz depths.
For larger subspace dimensions $K$,
convergence seems slower in the soft-ortho case compared to hard-ortho.
This effect is visible in the bottom-right panel of Figure~\ref{fig:cost-itovpars-ising},
where the soft-ortho cost curves ($K=8$) do not seem to reach a plateau
within the maximum number of iterations.

\begin{figure}
	\centering
	\includegraphics[width=1\linewidth]{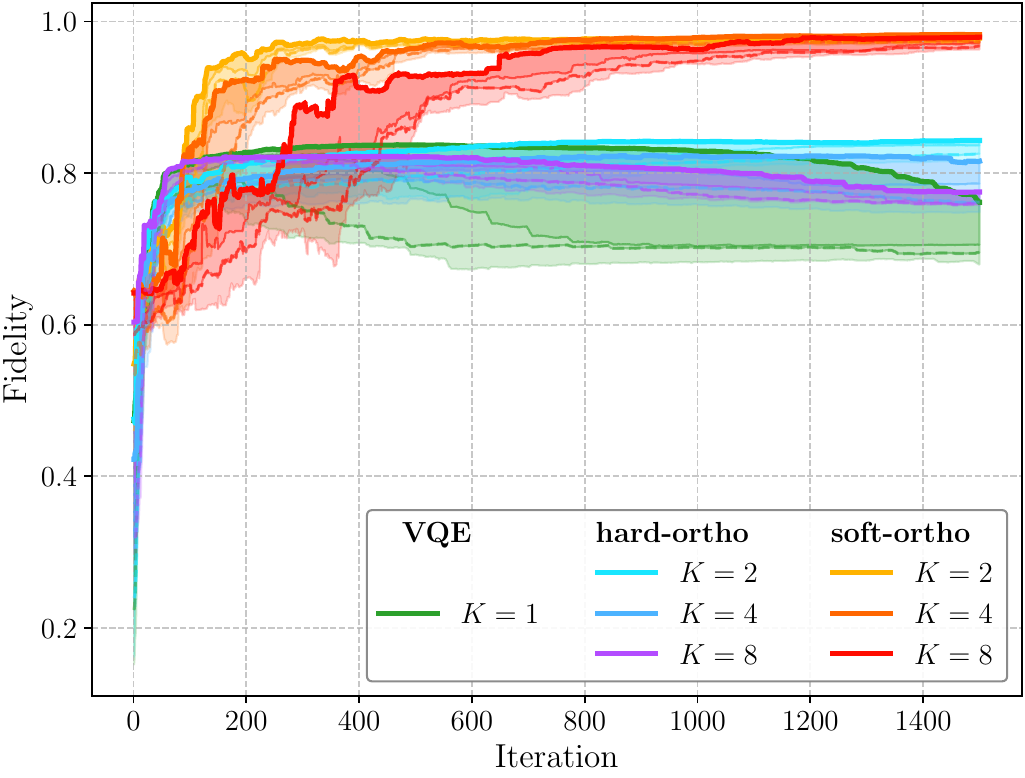}
	\caption{Behavior of the fidelity $\mathcal{F}_{\rm trc}$ between the exact
		ground state and the variational ground state candidate as a function of
		optimization steps for standard VQE ($K=1$), hard-ortho and soft-ortho subspace
		representations with subspace dimensions $K=2,4,8$ for the $3\times 3$ 
		transverse-field Ising model. All ansätze have $N_l=4$ layers as described
		in the text. Solid lines show the best fidelity across 10 independent
		runs at each iteration, dashed lines indicate the median, and shaded
		bands indicate the spread of the fidelity values from the best run up
		to the 25th, 50th and 75th percentiles (darker to lighter).}
	\label{fig:fiditising}
\end{figure}
As the subspace cost decreases, one would expect
a general increase in the fidelity of the trial ground state.
To verify this, we fix the number of layers to $N_l=4$ for all ans\"atze
and compare the behavior of the $H_{\rm trc}$ fidelity
$\mathcal{F}_{\rm trc}$ as a function of the number of optimization
iterations in Figure~\ref{fig:fiditising},
which involves 10 independent runs for each fixed setup.
The fidelity for standard VQE reaches at best approximately $76\%$,
while for hard-ortho it stabilizes around $80\%$ for all values 
of the subspace dimension $K$ investigated.
We observe a generally better performance of the soft-ortho representation,
with best-run fidelities above $97\%$ across all subspace dimensions.
The spread across independent runs, shown by the shaded bands,
is largest for VQE and smallest for both hard-ortho and soft-ortho.
We stress that, while the loss decreases with the number of iterations,
this does not always corresponds to a monotonic increase of the fidelity.
In general,
even if loss and fidelity are expected to be somewhat anti-correlated,
there is not guarantee that a lower cost corresponds to a higher fidelity,
unless the overlap with higher energy states can be considered negligible.
For example, the best fidelity of standard VQE runs shown in Figure~\ref{fig:fiditising}
exceed $80\%$ at intermediate iterations before dropping in later steps.

\begin{figure*}
	\centering
	\includegraphics[width=1\linewidth]{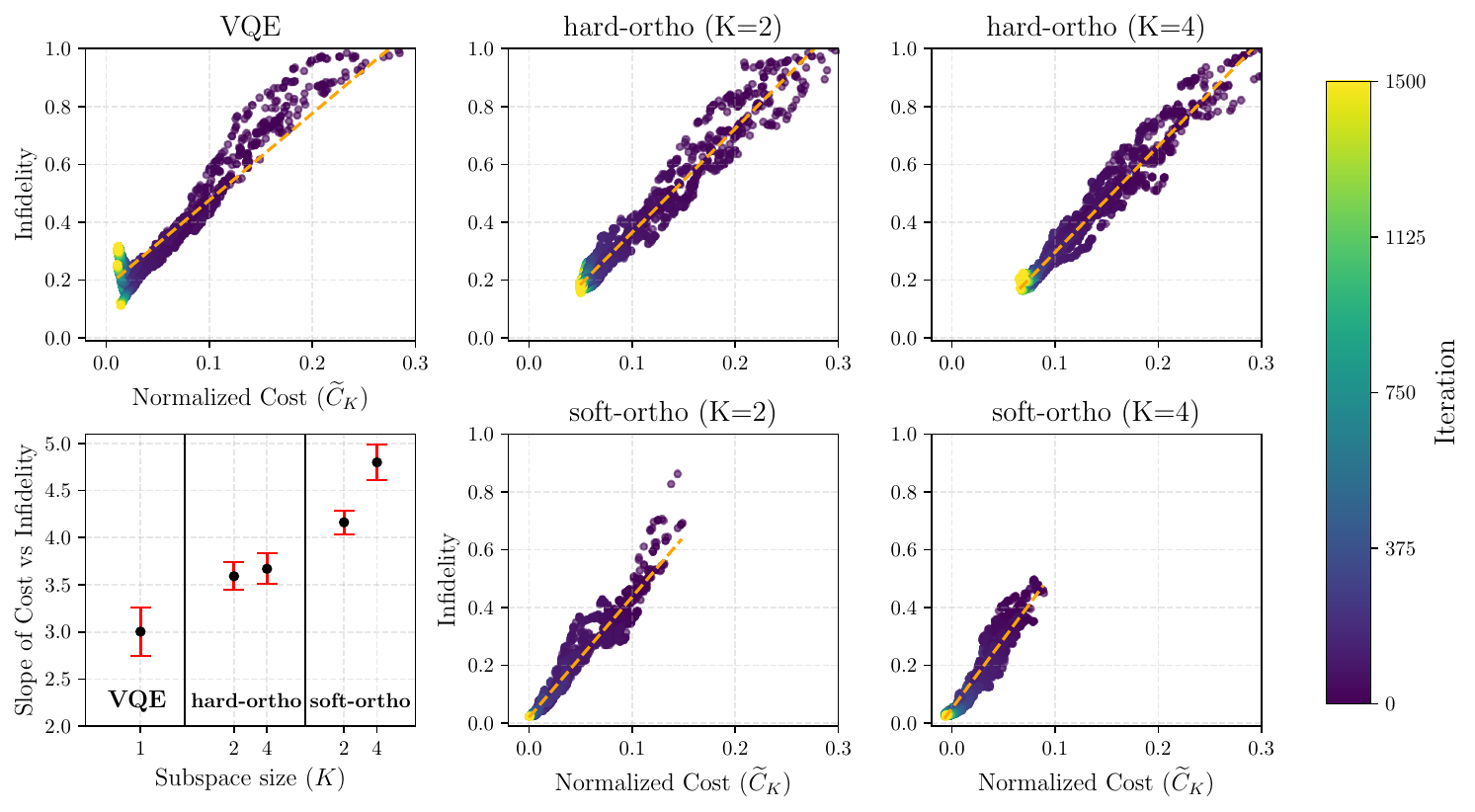}
	\caption{Scatter plot of the infidelity $1 - \mathcal{F}_{\rm trc}$ versus
		the normalized cost $\widetilde{C}_K$ (Eq.~\eqref{eq:normalized-cost})
		for standard VQE (top left), hard-ortho (top center and right, $K=2,4$) and
		soft-ortho (bottom center and right, $K=2,4$) representations on the
		$3 \times 3$ transverse-field Ising model. All ansätze have 144 trainable
		parameters. Each point corresponds to one optimizer iteration step in one
		of the 10 independent runs; the color indicates the iteration number (see
		colorbar). For each run, we fit a linear relation
		$1 - \mathcal{F}_{\rm trc} \simeq m \widetilde{C} + q$,
		showing the average slope $m$ of the fits across all 10 runs as a dashed orange line.
		The bottom-left panel shows these averages $m$ (with standard error for the errorbars).}
	\label{fig:correlation}
\end{figure*}
To investigate this aspect further, we consider the relation
between the infidelity $1 - \mathcal{F}_{\rm trc}$ and the value of the cost function
during the optimization.
To this end, we introduce a \emph{normalized cost} which,
for an exactly orthogonal frame, takes values between 0 and 1:
\begin{equation}
	\widetilde{C}_K := \frac{C_K - C_K^{\min}}{C_K^{\max} - C_K^{\min}},
	\label{eq:normalized-cost}
\end{equation}
where $C_K = \sum_{p=0}^{K-1} \mel{\psi_p}{H}{\psi_p}$
is the total frame energy,
while $C_K^{\min}$ and $C_K^{\max}$ are, respectively,
the energies of the $K$ lowest and highest exact eigenstates.
Notice that in this definition we exclude the $\beta$ penalty terms for the soft-ortho,
which are present in the actual cost function~\eqref{eq:cost-soft}
optimized during the run.
By construction, $\widetilde{C}_K = 0$ when the subspace overlaps exactly
with the $K$ lowest eigenstates.
Furthermore, in the soft-ortho case, the value of $\widetilde{C}_K $ may lie outside $[0,1]$,
since the states are not forced to be strictly orthogonal.
In Figure~\ref{fig:correlation} we show the relation between the normalized cost
$\widetilde{C}_K$ and the infidelity $1 - \mathcal{F}_{\rm trc}$ during optimization
for 10 independent runs, where all ans\"atze are chosen
so that the number of trainable parameters is fixed to 144 for a fairer comparison.
All representations show an approximately linear relation on the first part of
the optimization (iterations grows from darker to lighter in the colormap).
The bottom-left panel reports instead the average slopes over 10 runs for the
five cases considered in the other panels:
soft-ortho has a slope roughly $1.5$ times larger than VQE and hard-ortho.
This slope indicates how a reduction in cost estimates
corresponds to a larger improvement in fidelity (or decreasing infidelity).

As mentioned for Figure~\ref{fig:fiditising},
some runs of standard VQE (and, to a lesser extent, the hard-ortho case)
exhibit a non-monotonic behavior where the cost function decrease
is accompanied by an increase in infidelity.
This indicates that the cost function alone is not a good surrogate for fidelity in these cases,
especially in the latest stage of the optimization.
On the other hand, soft-ortho runs seem to follow more direct paths towards higher fidelities.
It is also interesting to notice that, while the initial infidelity values
of both the standard VQE and the hard-ortho runs are relatively large
(i.e., low overlap with the ground state),
most of the the runs in the soft-ortho representation
start with already reasonably low infidelities.
Regarding the pairwise overlap $\left|\braket{\psi_q(\bm \theta_{(q)})}{\psi_p(\bm \theta_{(p)})}\right|^2$
between distinct frame states from the soft-ortho representation,
we monitored their value during the optimization.
At the beginning of the run, the random initialization of the parameters
does not make it meaningful,
while at the end of the optimization, where the penalty term takes
effect over the relative energy gaps, we never measured values exceeding $15\%$
for all subspace dimensions $K$ and number of layers $N_l$ investigated.
Also, it can reach values up to $30\%$ during the run
(besides the very first few iterations).
Moreover, for $K=2$ the overlap between the two frame states,
while possibly peaking during the optimization up to $20\%$,
reaches overlaps lower than $5\%$ at the end of the run.

\begin{table}
	\centering
	\begin{tabular}{|c|c|c|c|c|c|}
		\hline
		\multicolumn{2}{|c|}{\textbf{}} & \multicolumn{2}{c|}{\textbf{hard-ortho}} & \multicolumn{2}{c|}{\textbf{soft-ortho}} \\
    \cline{1-6} {\scriptsize $K$} & {\scriptsize Layers ($N_l$)} & {\scriptsize $\max{\Delta {\cal F}}$ \%} & {\scriptsize ${\rm med}{\Delta {\cal F}}$ \%} & {\scriptsize $\max{\Delta {\cal F}}$ \%} & {\scriptsize ${\rm med}{\Delta {\cal F}}$ \%} \\
		\hline\hline
		\multirow{6}{*}{2} 
		& 2  & -    & -      & 0.3 & \textless 0.1 \\
		\cline{2-6} 
		& 4  & 0.7  & \textless 0.2 & 0.1 & 0.02(1) \\
		\cline{2-6} 
		& 6  & 1.0  & 0.5(2) & 0.6 & \textless 0.02 \\
		\cline{2-6} 
		& 8  & 1.3  & 0.5(1) & 0.2 & \textless 0.04 \\
		\cline{2-6} 
		& 10 & 1.1  & 0.8(2) & -   & -       \\
		\cline{2-6} 
		& 12 & 1.1  & \textless 0.3 & -   & -       \\
		\hline
		\multirow{6}{*}{4} 
		& 2  & -    & -       & 0.2 & 0.15(2) \\
		\cline{2-6} 
		& 4  & 0.9  & 0.73(5) & 0.3 & 0.10(2) \\
		\cline{2-6} 
		& 6  & 1.0  & 0.76(4) & 0.4 & 0.07(5) \\
		\cline{2-6} 
		& 8  & 0.8  & 0.70(5) & 0.5 & 0.13(5) \\
		\cline{2-6} 
		& 10 & 1.2  & 0.78(8) & -   & -       \\
		\cline{2-6} 
		& 12 & 0.9  & 0.74(6) & -   & -       \\
		\hline
		\multirow{4}{*}{8} 
		& 2  & -    & -       & 0.3 & 0.11(5) \\
		\cline{2-6} 
		& 4  & 1.3  & 1.25(1) & 0.1 & 0.07(1) \\
		\cline{2-6} 
		& 6  & 1.6  & 1.21(5) & 0.8 & 0.2(1)  \\
		\cline{2-6} 
		& 8  & 1.4  & 1.29(3) & 0.9 & 0.29(7) \\
		\hline
	\end{tabular}
	\caption{Maximal and median percentual discrepancies $\Delta {\cal F}={\cal F}_{\rm sub}-{\cal F}_{\rm trc}$
	between the subspace (Eq.~\eqref{eq:def-subfid}) and $H_{\rm trc}$ fidelity (Eq.~\eqref{eq:def-htrcfid})
	across 10 independent runs for each fixed number
	layers ($N_l$) or subspace dimension $K$ for the transverse-field Ising model on a $3\times 3$ lattice.
	Missing values are flagged with a hyphen and correspond to simulation not performed,
  while elements with `$\textless$' denote values compatible with errors comparable with the estimate, 
  so we report the upper bound of the confidence interval at 80\% confidence level.
}
	\label{tab:fidelity_L3}
\end{table}
Since the fidelity data reported in Figures~\ref{fig:fiditising} and~\ref{fig:correlation}
refer only to the $H_{\rm trc}$ definition, i.e., the fidelity of the state
obtained from the diagonalization of the truncated Hamiltonian,
one might wonder how this compares with the subspace fidelity $\mathcal F_{\rm sub}$,
which measures the overlap of the exact ground state with the subspace itself,
as introduced in Section~\ref{subsec:subspace-fidelities}.
To this end, in Table~\ref{tab:fidelity_L3} we report\footnote{
Errors in tables are estimated with bootstrap resampling for quantities involving medians, 
while quantities involving extremal values (e.g., min, max) are reported without error (see Ref.~\cite{Bickel1981,Angus01011992} for a discussion).}
the median and maximal discrepancy \mbox{$\Delta {\cal F} := \mathcal F_{\rm sub} - \mathcal F_{\rm trc}$}
at the end of the optimization runs
for both hard-ortho and soft-ortho representations,
computing it across the 10 independent runs for each ansatz configuration.
In all cases considered, the discrepancy never exceeds $2\%$ for the hard-ortho
and $1\%$ for the soft-ortho representation.
This observation confirms that, at least for the present model and ansatz structures,
the subspace at convergence ${\cal V}_{\bm \theta^*}$ is well aligned
with an eigensector of the Hamiltonian (possibly the low-energy one),
with relatively negligible mixing with
the complementary subspace $\overline{\cal V}_{\bm \theta^*}$.
Therefore, the best estimate obtained from the diagonalization of the truncated Hamiltonian
is not too far (within the percent level) from the best possible state
that can be obtained from the variational subspace itself.
\begin{figure*}
	\centering
	\includegraphics[width=1\linewidth]{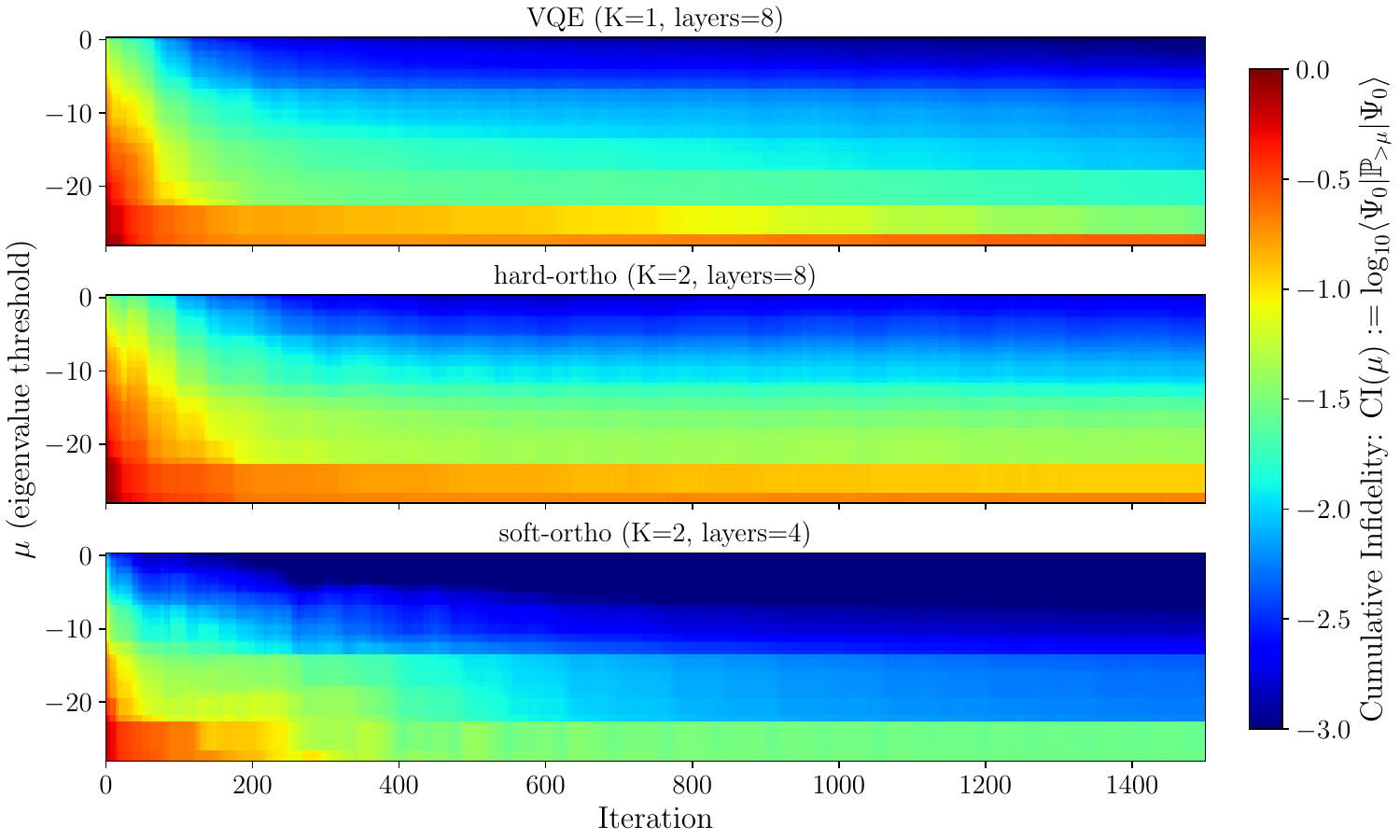}
	\caption{Color map of the cumulative infidelity ${\rm CI}(\mu)$
		(see Eq.~\eqref{eq:CI})
		during the optimization run for both VQE (top panel),
		hard-ortho (middle panel) and soft-ortho (bottom panel).
		All subspace ansätze have 144 trainable parameters 
    ($N_l=8$ layers for both VQE and hard-ortho, $N_l=4$ layers for soft-ortho). The color scale indicates the base-10 logarithmic overlap
		of the variational ground state candidate with exact eigenspaces with energy larger
		than the moving threshold $\mu$: darker blue regions indicate smaller contamination from
		high-energy states. }
	\label{fig:spectrograms-ising}
\end{figure*}
With this observation in mind,
we can further investigate how the ground state estimate $\ket{\Psi_0}$
overlaps with the low-energy and higher-energy eigensectors of the Hamiltonian $H$
during the optimization.
Indeed, merely considering the infidelity with respect to the exact ground state,
without discriminating between higher-energy contributions,
does not provide enough information about the quality of the variational state.
To obtain a more detailed picture of these sectors, we consider
a running energy threshold $\mu$
and introduce the \emph{cumulative infidelity} (CI) function as follows:
\begin{equation}\label{eq:CI}
	\begin{aligned}
		{\rm CI}(\mu) & := \log_{10}\mel{\Psi_0}{{\mathbb{P}}_{>\mu}}{\Psi_0}, 
	\end{aligned}
\end{equation}
where ${\mathbb{P}}_{>\mu}$ is the spectral projector
built from the eigenstates of $H$ with eigenvalues
higher than the threshold energy $\mu$, i.e.,
\begin{align}
	\mathbb{P}_{>\mu} := \sum_{\substack{\lambda > \mu \\ \lambda \in \sigma(H)}} \sum_{\alpha=0}^{d_\lambda -1}|{\phi_{\lambda,\alpha}^{\rm exact}}\rangle\!\langle{\phi_{\lambda,\alpha}^{\rm exact}}|.
\end{align}
Notice that, as $\mu$ crosses an eigenvalue $\lambda$ of $H$,
the full eigenspace of degeneracy $d_\lambda$ associated to it
is included in $\mathbb{P}_{\mu}$, excluding possible partial
and arbitrary overlaps with $\ket{\Psi_0}$, which could instead occur
if only a single eigenstate per eigenvalue were considered.

The behavior of the cumulative infidelity for the three approaches considered and
at a common fixed number of 144 parameters is shown in Figure~\ref{fig:spectrograms-ising}
for randomly picked individual runs\footnote{Threshold energies including the ground state $\mu \leq E_0$ are excluded from Figure~\ref{fig:spectrograms-ising},
	since ${\mathbb P}_{\mu \geq E_0}=\mathds{1}$ and they trivially correspond to ${\rm CI} = 0$.}.
At first, all state representations have a significant overlap with high-energy eigenstates,
while they progressively reduce it in favor of low-energy contributions as the optimization proceeds.
The standard VQE and hard-ortho representation show a similar behavior,
with significant overlaps with mid-energy eigenspaces even at the end of the optimization.
It is also worth looking,
especially for the standard VQE panel,
the very bottom side strip of the spectrum involving
the first excited eigenspace above the ground state.
Here we see that, as the overlap with higher-energy states continues to deplete
as the optimization run progresses,
the overlap including the first excited eigenspace shows a non-monotonic behavior:
up to intermediate iterations (around $\sim 600$) it lowers but,
going further in the optimization (i.e., iterations $\gtrsim 1000$)
the contamination from the first excited eigenspace increases again.
This indicates that, while optimizing the average energy  of the variational state,
the overall overlap with high-energy states is reduced but,
depending on the expressivity of the ansatz and accessible regions of the parameter space,
the overlap with the above-ground eigenspaces is not prevented from increasing,
especially at convergence where energy variations are smaller
and the dynamics of the optimization is more constrained 
by the ansatz structure.
This is in line with the non-monotonic behavior of the fidelity
observed in Figure~\ref{fig:fiditising} for standard VQE.
On the other hand, the soft-ortho representation
shows a more effective suppression not only of high-energy contributions,
but also of mid-energy and low-lying excited states contributions,
leading to a more robust convergence to the ground state.
Indeed, realizing that this very non-monotonic behavior of the ground state fidelity
is at play clarifies also why approaches involving multiple states
might be more effective even in tasks targeting only the ground state.
The hard-ortho approach however, while showing some improvements over standard VQE,
and mostly preventing this behavior, is still not as effective as the soft-ortho one
in reaching high fidelities,
possibly due to the more stringent constraints on the expressivity of the variational frame,
as argued in Section~\ref{subsec:subspace-express}.

\subsubsection{Fidelity Dependence on Subspace Dimension and Number of Parameters}\label{subsubsec:numres-fidelitysummary}

\begin{figure}
	\centering
	\includegraphics[width=0.97\linewidth]{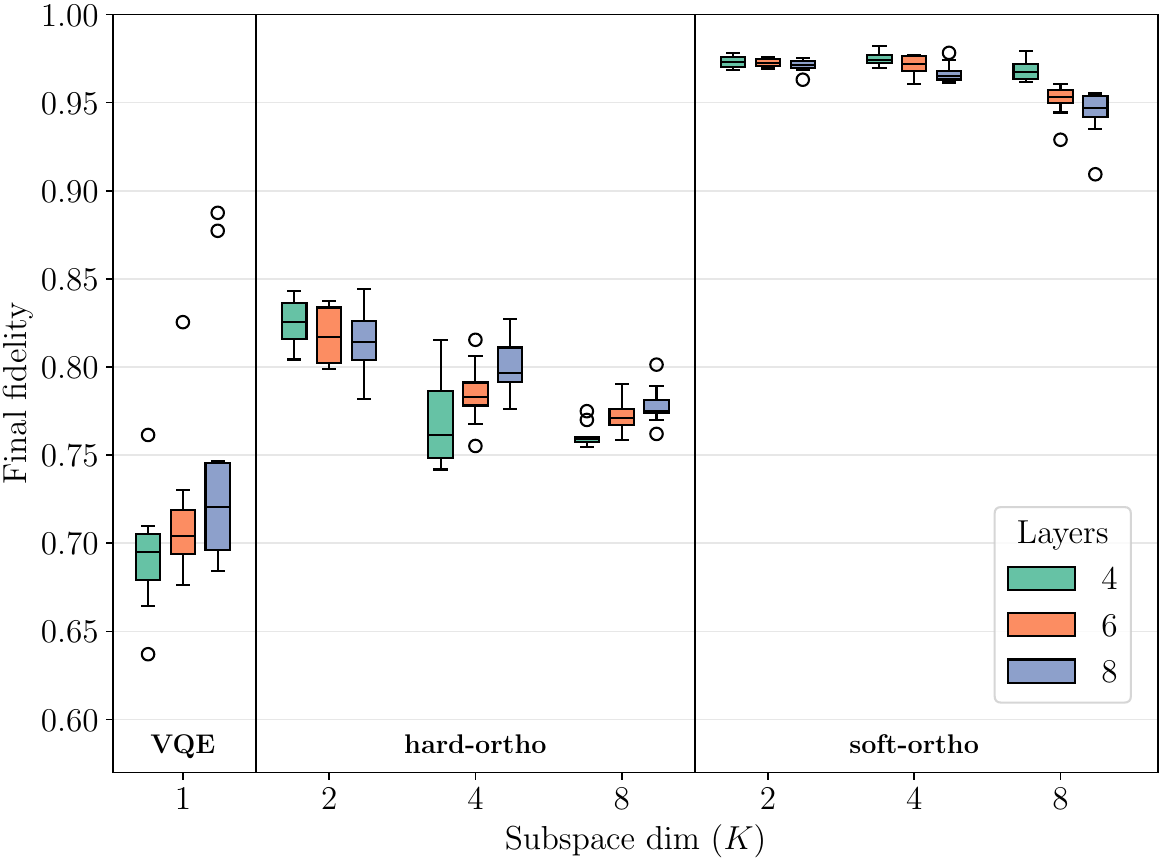}
	\caption{Box plot of the final ground state fidelities $\mathcal{F}_{\rm trc}$
		(after 1500 optimization steps) versus subspace dimension $K$ for standard
		VQE ($K=1$), hard-ortho and soft-ortho representations on the $3 \times 3$
		transverse-field Ising model. Colors indicate the number of ansatz layers
		($N_l=4$, 6 or 8). Each box represents the distribution of fidelities over 10 independent runs
		(line: median; box: interquartile range; whiskers: 1.5 x interquartile range;
		circles: outliers). Vertical lines separate the three algorithms.}
	\label{fig:fiditising_boxplot}
\end{figure}

\begin{figure}
	\centering
	\includegraphics[width=0.97\linewidth]{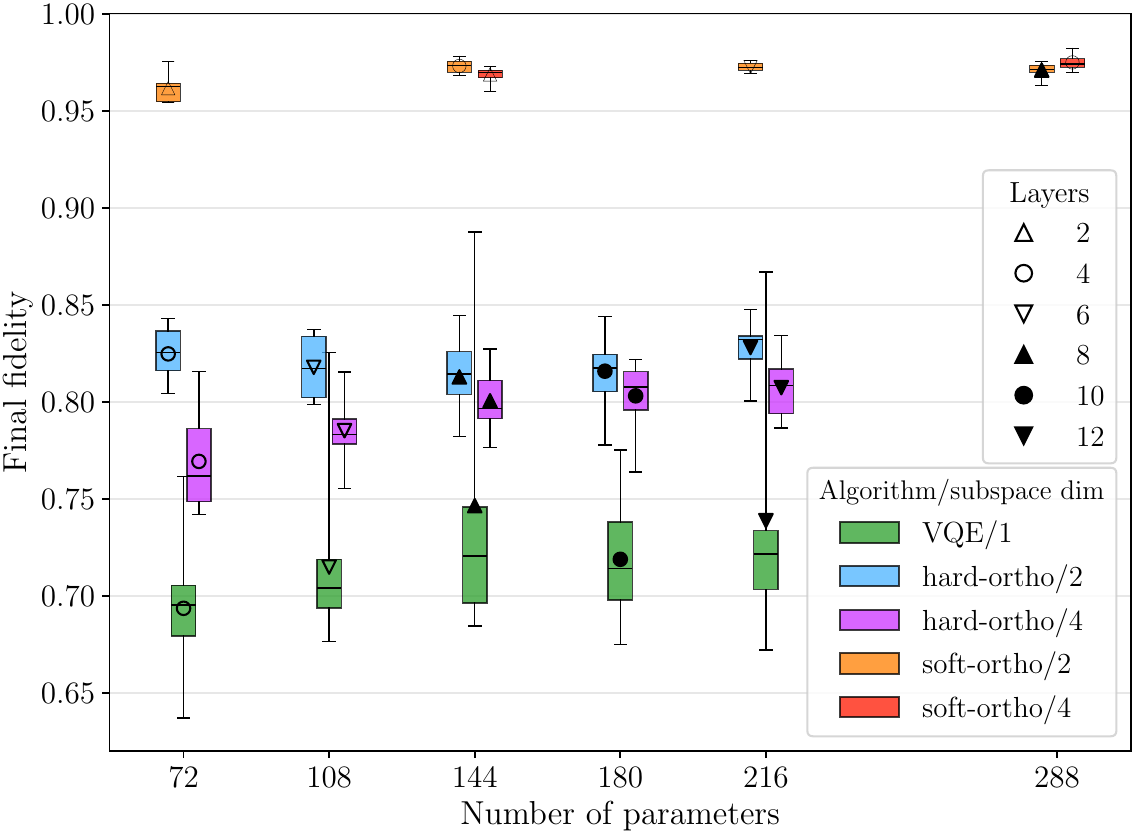}
	\caption{Box plot of the final ground state fidelities $\mathcal{F}_{\rm trc}$
		(after 1500 optimization steps) versus the number of trainable parameters
		for standard VQE, hard-ortho and soft-ortho
		representations (only for subspace dimensions $K=2$ and $K=4$)
		on the $3 \times 3$ transverse-field Ising model.
		Each box summarizes the distribution of fidelities over 10 independent runs
		(line: median; box: interquartile range; whiskers: full range;
		marker: mean). For clarity, boxes at the same number of parameters
		are slightly shifted horizontally.}
	\label{fig:fidvsparams_ising}
\end{figure}

\begin{table}
	\centering
		\begin{tabular}{|c|c|c|c|c|c|c|c|}
			\hline
			\multicolumn{2}{|c|}{\textbf{}}                  & \multicolumn{2}{c|}{\textbf{VQE}}            & \multicolumn{2}{c|}{\textbf{hard-ortho}}    & \multicolumn{2}{c|}{\textbf{soft-ortho}} \\
      \cline{1-8} K                   & $N_l$    & $\max {\cal F}$  & ${\rm med} {\cal F}$ & $\max {\cal F}$ & ${\rm med} {\cal F}$ & $\max {\cal F}$ & ${\rm med} {\cal F}$ \\ \hline\hline
			\multirow{6}{*}{2}              & 2     & -              & -                  & -             & -                  & 0.976         & 0.965(2)   \\
			\cline{2-8}                     & 4     & 0.76           & 0.70(1)            & 0.84          & 0.83(1)            & 0.978         & 0.973(2)   \\
			\cline{2-8}                     & 6     & 0.82           & 0.70(1)            & 0.84          & 0.82(1)            & 0.976         & 0.972(1)   \\
			\cline{2-8}                     & 8     & 0.89           & 0.71(2)            & 0.84          & 0.82(1)            & 0.976         & 0.971(1)   \\
			\cline{2-8}                     & 10    & 0.78           & 0.71(1)            & 0.84          & 0.82(1)            & -             & -          \\
			\cline{2-8}                     & 12    & 0.87           & 0.72(2)            & 0.85          & 0.84(1)            & -             & -          \\ \hline
			\multirow{6}{*}{4}              & 2     & -              & -                  & -             & -                  & 0.973         & 0.970(1)   \\
			\cline{2-8}                     & 4     & 0.76           & 0.70(1)            & 0.82          & 0.76(1)            & 0.982         & 0.974(1)   \\
			\cline{2-8}                     & 6     & 0.82           & 0.70(1)            & 0.82          & 0.78(1)            & 0.977         & 0.972(2)   \\
			\cline{2-8}                     & 8     & 0.89           & 0.71(2)            & 0.83          & 0.79(1)            & 0.978         & 0.964(2)   \\
			\cline{2-8}                     & 10    & 0.78           & 0.71(1)            & 0.82          & 0.81(1)            & -             & -          \\
			\cline{2-8}                     & 12    & 0.87           & 0.72(2)            & 0.83          & 0.81(1)            & -             & -          \\ \hline
			\multirow{4}{*}{8}              & 2     & -              & -                  & -             & -                  & 0.979         & 0.972(2)   \\
			\cline{2-8}                     & 4     & 0.76           & 0.70(1)            & 0.78          & 0.76(1)            & 0.979         & 0.968(2)   \\
			\cline{2-8}                     & 6     & 0.82           & 0.70(1)            & 0.79          & 0.77(1)            & 0.960         & 0.953(2)   \\
			\cline{2-8}                     & 8     & 0.89           & 0.71(2)            & 0.80          & 0.77(1)            & 0.955         & 0.947(3)   \\ \hline
		\end{tabular}
	\caption{Comparison of best and median fidelities $\mathcal{F}_{\rm trc}$
		after 1500 iterations for standard VQE, hard-ortho and soft-ortho
		representations and different ansatz layers ($N_l$) 
    and subspace dimensions ($K$)
		for the $3\times 3$ transverse-field Ising model considered in this section.
		The VQE results are independent of $K$ and are duplicated to ease
		comparison with the subspace approaches. 
  Missing values are flagged with a hyphen and correspond to simulation not performed.}
	\label{tab:fidelity_full_L3}
\end{table}
A summary of results for the final best and median
$H_{\rm trc}$ fidelities (after 1500 iterations)
for the three approaches considered and different ansatz depths and subspace dimension $K$
is reported in Table~\ref{tab:fidelity_full_L3}, where the general
behavior discussed above is confirmed.
The cases with larger subspace dimension $K$ seem to be more challenging
for both hard-ortho and soft-ortho representations, with a slight decrease
in the best fidelities achieved at fixed number of layers,
which seems to indicate a generally slower convergence.
To get a better picture of the general dependence
of the final fidelity, we provide two points of views:
one relating the final fidelity ${\cal F}_{\rm trc}$
with the subspace dimension $K$ (Figure~\ref{fig:fiditising_boxplot}),
and another one relating it with the number of trainable parameters in the (subspace) ansatz (Figure~\ref{fig:fidvsparams_ising}).
Both perspectives reflect the distribution of final fidelities over 10 independent runs,
which are aggregated in the aforementioned Table~\ref{tab:fidelity_full_L3},
reporting only the maximum and the median fidelities.
Let us discuss first Figure~\ref{fig:fiditising_boxplot},
where we show the dependence of the distribution of final $\mathcal{F}_{\rm trc}$ fidelities
over 10 independent runs and as a function of the subspace dimension $K$.
Looking at the best fidelities achieved,
we can see that both VQE and hard-ortho remain in the $75$-$90\%$ range,
while soft-ortho reaches fidelities above $95\%$ for all values of $K$
and all numbers of layers in the ans\"atze.
In particular, hard-ortho shows no improvement over standard VQE
if looking at best runs, while a marginal improvement ($\sim 70\%$ to $\sim 80\%$)
is observed when looking at median fidelities.
For soft-ortho, we also see that increasing the subspace dimension beyond $K=2$
does not seem to improve the fidelity, or at least a possible marginal improvement
is traded for a slower convergence which makes the cut-off of 1500 iterations insufficient.
This is also evident in the hard-ortho case, even if its number of parameters
at fixed number of layers is the same as standard VQE.
From this observation, one might conclude that the advantage of a subspace
approach, at least for the present model and ansatz structures,
is mostly granted by the inclusion of just one additional state in the frame,
while further increasing the subspace dimension does not provide significant benefits.
This might be related to the fact that the first excited state
is the one mostly contaminating the variational ground state estimate
at the end of the optimization, as argued in Section~\ref{subsubsec:numres-TFI2D-lossfid}
for Figure~\ref{fig:spectrograms-ising};
including a second state to the variational frame seems to be already sufficient
to make the optimal subspace ${\cal V}_{\bm \theta^*}$ significantly overlap with the exact first excited state,
thus allowing in the diagonalization stage to discriminate ground state components with better accuracy.
The second perspective we provide on the final fidelities achieved is reported
in Figure~\ref{fig:fidvsparams_ising} with an emphasis
on the fixed number of trainable parameters.
Soft-ortho clearly outperforms standard VQE and hard-ortho
for all numbers of parameters explored and all runs.
While standard VQE and hard-ortho approaches show
a relatively large spread in the final fidelities across runs,
the final estimates obtained with soft-ortho are more robustly clustered
around high-fidelity values, with significantly smaller interquartile ranges,
confirming a more reliable convergence behavior.
Moreover, the worst run for the soft-ortho representation with only 72 parameters
(i.e., $N_l=2$ layers and $K=2$) already exceeds the best fidelity obtained in both VQE
and hard-ortho with all numbers of parameters considered.
This confirms that shallower circuits are already sufficient
to reach high accuracies when using the soft-ortho subspace representation.
Indeed, as observed for Figure~\ref{fig:fiditising_boxplot},
deeper circuits with larger number of parameters seem to improve the
median final fidelity just marginally and up to a point (around 144 parameters or $N_l=8$ layers),
after which it becomes counterproductive,
likely due to the cut-off imposed of maximum 1500 total iterations
(as evident from the cost histories in Figure~\ref{fig:cost-itovpars-ising}).
In general, as argued in Section~\ref{subsec:subspace-express},
we observe that sacrificing some entanglement expressivity
is not expected to quicken convergence but can be beneficial
on reaching higher accuracies:
frames with shallower circuits would reach lower levels of global entanglement,
but considering general linear combinations of independent states in the frame
allows to recover higher accuracies on the ground state estimates.

\subsubsection{Relative Gain Factor}\label{subsubsec:numres-relative-gain}
In this Section, we introduce and discuss a quantitative metric
to assess the improvement in accuracy achieved by
the subspace approaches with respect to standard VQE.
At first look, one might be tempted to directly
use the difference \mbox{$\delta := \max_r\mathcal F_{r, {\rm Algo}}-\max_{r'}\mathcal F_{r', {\rm VQE}}$}
between the best fidelities (or the median ones, if we replace $\max$ with ${\rm med}$)
achieved among different runs $r, r'$ for the same system at fixed number of parameters.
However, this quantity would value the incremental improvement irrespectively of
the absolute value of the fidelities achieved.
For example, a $\delta=10\%$ gain
obtained when improving the fidelity from $80\%$ to $90\%$ would be considered better than an
improvement of $\delta=8\%$ when the fidelity raises from $90\%$ to $98\%$, 
while intuition suggests the opposite.
For this reason, we introduce an empirical measure
of the relative gain of a subspace approach with respect to the standard VQE approach,
which we call \emph{relative gain factor} and for which we propose two versions:
\begin{align}
	\mathcal{G}_{\rm Algo}^{({\rm med})} = \frac{\underset{r}{\rm med}\{1-\mathcal F_{r, \rm VQE}\}}{\underset{r'}{\rm med}\{1-\mathcal F_{r', \rm Algo}\}}, \label{eq:def-gain-med} \\
	\mathcal{G}_{\rm Algo}^{({\rm min})} = \frac{\underset{r}{\rm min}\{1-\mathcal F_{r, \rm VQE}\}}{\underset{r'}{\rm min}\{1-\mathcal F_{r', \rm Algo}\}}, \label{eq:def-gain-min}
\end{align}
where ${\rm min}/{\rm med}$ are aggregate functions denoting the minimum or median
of the infidelity $1-\mathcal F_{r, {\rm Algo}}$ over the different runs $r$ performed for each algorithm
\mbox{${\rm Algo}\in \{\text{hard-ortho}, \text{soft-ortho}\}$} on the same system.
Thus, ${\cal G}$ measures the ratio between the smallest/median infidelity
among standard VQE runs with respect to the same quantity obtained
with the hard-ortho or soft-ortho subspace representation.
A gain factor of $\mathcal{G}=1$ indicates no relative improvement (by definition, $\mathcal{G}_{VQE}=1$),
while values $\mathcal{G}>1$ indicate a proper improvement.
In the example mentioned above, the gain factor in the first case would be $\mathcal{G}=2$ (since the infidelity is halved),
while in the second case it would be $\mathcal{G}=5$, indicating a greater relative improvement,
as we intended.
\begin{table} 
	\centering
	\begin{tabular}{|c|c|c|c|c|c|}
		\hline
		\multicolumn{2}{|c|}{\textbf{}}        & \multicolumn{2}{c|}{\textbf{hard-ortho}} & \multicolumn{2}{c|}{\textbf{soft-ortho}} \\
		\cline{1-6} $K$      & $N_l$       & $\mathcal{G}_{\rm med}$ & $\mathcal{G}_{\min}$          & $\mathcal{G}_{\rm med}$ & $\mathcal{G}_{\min}$          \\ \hline\hline                                                                                                                                                                                                 
		\multirow{5}{*}{2} & 4        & 1.76(9)     & 1.52              & 11.4(9)     & 10.97             \\
		\cline{2-6}        & 6        & 1.63(11)    & 1.07              & 10.8(6)     & 7.28              \\
		\cline{2-6}        & 8        & 1.50(12)    & 0.72              & 9.7(8)      & 4.60              \\
		\cline{2-6}        & 10       & 1.55(9)     & 1.44              & -           & -                 \\
    \cline{2-6}        & 12       & 1.67(7)     & 0.87              & -           & -                 \\ \hline                                                                                                                                                                                                 
		\multirow{5}{*}{4} & 4        & 1.30(7)     & 1.29              & 12.1(9)     & 13.55             \\
		\cline{2-6}        & 6        & 1.37(4)     & 0.95              & 11(1)       & 7.68              \\
		\cline{2-6}        & 8        & 1.38(10)    & 0.65              & 8.0(7)      & 5.18              \\
		\cline{2-6}        & 10       & 1.47(8)     & 1.26              & -           & -                 \\
		\cline{2-6}        & 12       & 1.46(8)     & 0.80              & -           & -                 \\ \hline                                                                                                                                                                                                 
		\multirow{3}{*}{8} & 4        & 1.27(3)     & 1.06              & 9.4(8)      & 11.50             \\
		\cline{2-6}        & 6        & 1.29(4)     & 0.83              & 6.3(4)      & 4.42              \\
		\cline{2-6}        & 8        & 1.25(9)     & 0.57              & 5.3(6)      & 2.51              \\ \hline
	\end{tabular}
	\caption{Relative gain factor of the $H_{\rm trc}$ infidelity of subspace representations
	with respect to the standard VQE with median-vs-median ($\mathcal{G}_{\text{med}}$) 
   and best-vs-best ($\mathcal{G}_{\min}$) computed with 10 independent run 
   and for different values of the subspace dimension $K$ and number of ansatz layers $N_l$
  for the $3\times 3$ transverse-field Ising model considered in this Section.
	Missing values are flagged with a hyphen and correspond to simulation not performed.}
	\label{tab:gain_L3}
\end{table}
The relative gain for the transverse-field Ising model considered here
can be inferred from the best or median fidelity data reported in Table~\ref{tab:fidelity_full_L3},
and from the results shown in Figures~\ref{fig:fiditising_boxplot}
and~\ref{fig:fidvsparams_ising},
but, for the sake of clarity, we explicitly report them in Table~\ref{tab:gain_L3}.
We can see that the hard-ortho representation does indeed provide
an advantage over standard VQE when looking at median fidelities,
but can be even counterproductive when looking at the best runs,
since the best final fidelity of standard VQE attained can outperform
the best ones from the hard-ortho case.
The situation is quite different for the soft-ortho representation,
which systematically provides a significant relative gain which,
at least at $K=2$, which we identified as the most effective subspace dimension
for the present model, lattice sizes and ansatz structure,
can reach values above $10$ for median cases
and above $4$ even for the best runs.

\subsection{Results for Edwards--Anderson Model}\label{subsec:numres-spin-glass}

	As a more challenging testbed for the subspace approaches investigated,
	in this Section we consider the 2D Edwards--Anderson spin glass model,
	described in the introduction to Section~\ref{sec:numres}.
	This model follows the general Ising-like Hamiltonian in Eq.~\eqref{eq:IsinglikeHam}.
	with disordered couplings $J_{ij}$ sampled from independent gaussian distributions
	with zero mean and unit variance and a uniform transverse field fixed to $h=2$.
	Every sampling of all the couplings $J_{ij}$ attached to the edges of the $4\times 4$ lattice
	defines a different \emph{realization} for the spin glass model.
	In our numerical investigation, we emulated 10 independent realizations,
	performing 8 independent optimization runs each,
	with random initializations of the ansatz parameters in the range $[-0.2\pi, 0.2\pi]$
	(same choice as for the TFI model).
	Building on the insights gained from the TFI model in Section~\ref{subsec:numres-TFI2D},
	and for the sake of computational accessibility,
	in this case we focus on a single subspace ansatz configuration, with
	subspace dimension $K=2$ and $N_l=4$ layers for both hard-ortho and soft-ortho.
	This choice already provides significant improvements over standard VQE in the TFI case
	and is found to be reasonably effective also for the EA model (as confirmed below, a posteriori).
	For the soft-ortho case, we set $\beta = 2.5$,
	which fulfills weak pairwise orthogonality of the frame at convergence
	in all the realizations considered.

\begin{figure}
	\centering
	\includegraphics[width=1\linewidth]{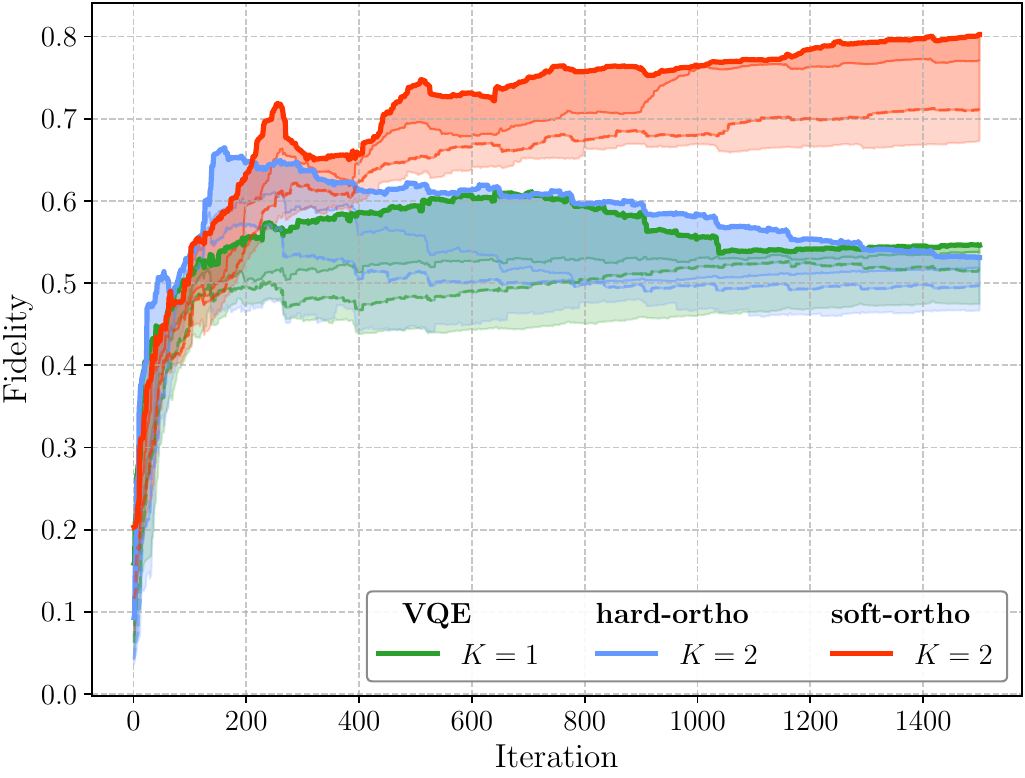}
	\caption{Fidelity $\mathcal{F}_{\rm trc}$ dependence during optimization
		between the exact ground state and the variational ground state candidate
		for standard VQE ($K=1$), hard-ortho and soft-ortho subspace
		representations with subspace dimension $K=2$ and for a selected realization
		(realization `10' in Tables~\ref{tab:Glass_Instance_Fid}
		and~\ref{tab:Glass_Instance_Gain})
		of the Edwards--Anderson model on a $4\times 4$ square lattice.
		All ansätze have $N_l=4$ layers as described in the text. 
    Solid lines show the best fidelity across 8 independent
		runs at each iteration step, dashed lines indicate the median, and shaded
		bands indicate the spread of the fidelity values from the best run up
		to the 25th, 50th and 75th percentiles (darker to lighter).}
	\label{fig:fiditglass}
\end{figure}
Let us consider first the evolution of the fidelity
	$\mathcal{F}_{\rm trc}$ (see definition in Eq.~\eqref{eq:def-htrcfid}) 
	during the optimization for one selected realization
	of the disordered couplings $J_{ij}$ in Figure~\ref{fig:fiditglass}.
	Here we observe a non-monotonic behavior of the fidelity with the number of iterations
	for both standard VQE and hard-ortho approaches,
	which is even pronounced than the one already discussed 
  for the TFI model in Figure~\ref{fig:fiditising}.
	The soft-ortho representation, while showing a larger spread
	across different runs with respect to the TFI case,
	exhibits instead a more consistent behavior of the fidelities,
	even if a higher number of iterations (above 1500) 
  could improve even further the final fidelity.

\begin{table} 
	\centering
	\begin{tabular}{|c|c|c|c|c|c|}
		\hline
		\multirow{2}{*}{\textbf{real.}}
		            & \multicolumn{2}{c|}{\textbf{hard-ortho}} & \multicolumn{2}{c|}{\textbf{soft-ortho}}                                  \\
		\cline{2-5} & {\scriptsize $\max{\Delta {\cal F}}$ \%} & {\scriptsize ${\rm med}{\Delta {\cal F}}$ \%} & {\scriptsize $\max{\Delta {\cal F}}$ \%} & {\scriptsize ${\rm med}{\Delta {\cal F}}$ \%} \\ \hline\hline
		1           & 0.446         & 0.25(5)         & 8.50     & 3.9(17)  \\
		2           & 1.923         & 0.55(37)        & 1.14     & 0.31(22) \\
		3           & 1.268         & 0.90(10)        & 3.71     & 0.98(38) \\
		4           & 0.417         & 0.31(6)         & 8.00     & 2.6(14)  \\
		5           & 1.026         & 0.57(27)        & 5.43     & 1.76(75) \\
		6           & 0.484         & 0.30(5)         & 7.16     & \textless 1.5   \\
		7           & 0.295         & \textless 0.08         & 0.52     & 0.11(7)  \\
		8           & 0.054         & 0.025(7)        & 2.20     & 0.80(48) \\
		9           & 0.336         & 0.20(6)         & 5.41     & \textless 2.2  \\
		10          & 0.446         & 0.20(9)         & 0.49     & 0.19(9)  \\ \hline
	\end{tabular}
	\caption{Maximal and median percentual discrepancies  $\Delta {\cal F}={\cal F}_{\rm sub}-{\cal F}_{\rm trc}$
	between the subspace (Eq.~\eqref{eq:def-subfid}) and $H_{\rm trc}$ fidelity (Eq.~\eqref{eq:def-htrcfid})
	across 8 independent runs for each 10 realizations
	of the Edwards--Anderson spin glass model on a $4\times 4$ lattice.
	The number of layers is fixed to $N_l=4$, while the subspace dimension is $K=2$.
  Elements with `$\textless$' denote values compatible with errors comparable with the estimate, 
  so we report the upper bound of the confidence interval at 80\% confidence level.}
	\label{tab:fidelity_gaps_Ising_new}
\end{table}
As with the TFI model, we also monitored the
	discrepancies between the subspace fidelity $\mathcal{F}_{\rm sub}$
	and the $H_{\rm trc}$ fidelity $\mathcal{F}_{\rm trc}$,
	reporting relevant aggregated data in Table~\ref{tab:fidelity_gaps_Ising_new}.
	For some realizations, we find somewhat larger discrepancies,
	possibly due to the increased complexity
	of the EA model with respect to the TFI one.
\begin{table}[t]
	\centering
	\begin{tabular}{|c|cc|cc|cc|}
		\hline
		\multirow{2}{*}{\textbf{real.}} & \multicolumn{2}{c|}{\textbf{VQE}} & \multicolumn{2}{c|}{\textbf{hard-ortho}} & \multicolumn{2}{c|}{\textbf{soft-ortho}}                                                           \\
		\cline{2-7}                     & ${\rm med}{\cal F}$               & $\max {\cal F}$                          & ${\rm med}{\cal F}$                      & $\max {\cal F}$ & ${\rm med}{\cal F}$ & $\max {\cal F}$ \\ \hline\hline
		1                               & 0.47(1)                         & 0.477                                  & 0.459(3)                                 & 0.489           & 0.79(4)             & 0.879           \\
		2                               & 0.51(1)                         & 0.570                                  & 0.53(3)                                  & 0.608           & 0.80(5)             & 0.794           \\
		3                               & 0.48(1)                         & 0.512                                  & 0.51(2)                                  & 0.575           & 0.84(3)             & 0.852           \\
		4                               & 0.39(1)                         & 0.512                                  & 0.38(1)                                  & 0.417           & 0.72(5)             & 0.759           \\
		5                               & 0.47(2)                         & 0.691                                  & 0.59(4)                                  & 0.717           & 0.85(3)             & 0.827           \\
		6                               & 0.50(1)                         & 0.568                                  & 0.47(1)                                  & 0.515           & 0.86(2)             & 0.868           \\
		7                               & 0.70(3)                         & 0.866                                  & 0.69(2)                                  & 0.775           & 0.88(1)             & 0.909           \\
		8                               & 0.50(7)                         & 0.518                                  & 0.477(3)                                 & 0.498           & 0.92(5)             & 0.920           \\
		9                               & 0.48(1)                         & 0.502                                  & 0.45(1)                                  & 0.495           & 0.72(6)             & 0.828           \\
		10                              & 0.53(2)                         & 0.546                                  & 0.50(2)                                  & 0.531           & 0.70(3)             & 0.802           \\ \hline
	\end{tabular}
	\caption{Final median and best fidelities (${\cal F}_{\rm trc}$, defined in Eq.~\eqref{eq:def-htrcfid})
		over 8 independent runs for each of the 10 realizations considered of the Edwards--Anderson model.}
	\label{tab:Glass_Instance_Fid}
\end{table}
	However, even in these cases, looking at the final ${\cal F}_{\rm trc}$ fidelities
  in Table~\ref{tab:Glass_Instance_Fid},
	we still observe a significant improvement of the soft-ortho approach
	with respect to both the standard VQE and the hard-ortho one.

\begin{figure}
	\centering
	\includegraphics[width=1.0\linewidth]{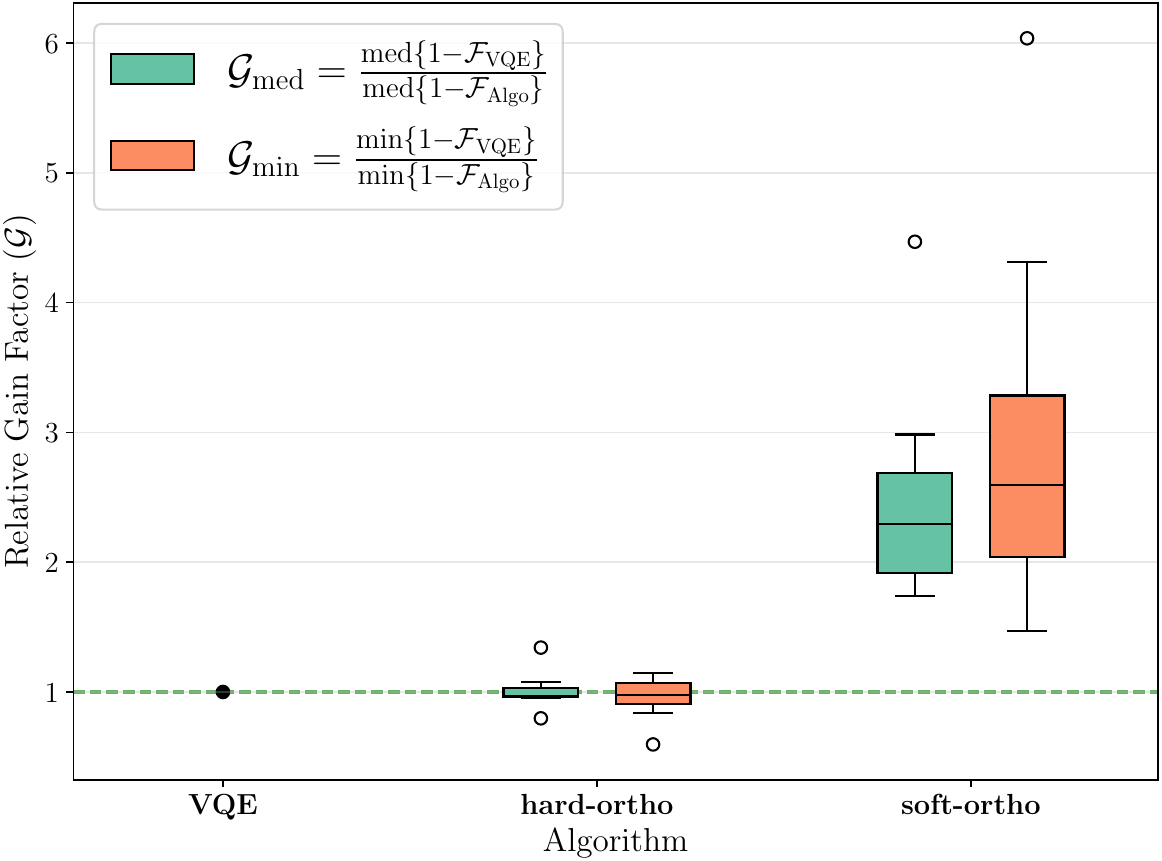}
	\caption{Box plot of the relative gain factor $\mathcal{G}$ in ground state infidelity
		with respect to standard VQE (see text for definition)
		for the 10 disorder realizations of the Edwards--Anderson spin-glass model on a
		$4 \times 4$ lattice. Green boxes show $\mathcal{G}_{\rm med}$ (median-based),
		orange boxes show $\mathcal{G}_{\rm min}$ (best-run-based).
		By definition, ${\cal G} = 1$ for VQE (dashed line).
		Values ${\cal G} > 1$ indicate improvement over standard VQE.
		The soft-ortho representation shows a clear gain,
		while hard-ortho performs similarly to VQE.}
	\label{fig:EA-relgain}
\end{figure}
\begin{table}[t]
	\centering
	\begin{tabular}{|c|cc|cc|}
		\hline
		\multirow{2}{*}{\textbf{real.}} & \multicolumn{2}{c|}{\textbf{hard-ortho}} & \multicolumn{2}{c|}{\textbf{soft-ortho}}                                            \\
		\cline{2-5}            & ${\cal G}_{\rm med}$              & ${\cal G}_{\min}$                        & ${\cal G}_{\rm med}$ & ${\cal G}_{\min}$ \\ \hline\hline
		1                      & 0.99(1)                         & 1.02                                     & 2.4(3)               & 4.31              \\
		2                      & 1.08(8)                         & 1.10                                     & 1.9(3)               & 2.09              \\
		3                      & 1.04(3)                         & 1.15                                     & 2.6(4)               & 3.29              \\
		4                      & 0.96(2)                         & 0.84                                     & 1.9(3)               & 2.03              \\
		5                      & 1.3(1)                          & 1.09                                     & 2.6(4)               & 1.79              \\
		6                      & 0.95(3)                         & 0.89                                     & 3.0(3)               & 3.28              \\
		7                      & 0.8(1)                          & 0.60                                     & 2.1(3)               & 1.47              \\
		8                      & 0.97(1)                         & 0.96                                     & 4.5(14)              & 6.04              \\
		9                      & 0.96(2)                         & 0.98                                     & 1.8(5)               & 2.90              \\
		10                     & 0.97(5)                         & 0.97                                     & 1.7(2)               & 2.30              \\ \hline
	\end{tabular}
	\caption{Final gains (both $\mathcal{G}_{\rm med}$ and $\mathcal{G}_{\min}$,
		as defined by Eqs.~\eqref{eq:def-gain-med} and~\eqref{eq:def-gain-min})
		over 8 independent runs for each of the 10 realizations considered
		of the Edwards--Anderson model. }
	\label{tab:Glass_Instance_Gain}
\end{table}
  To better quantify the improvement over VQE across different disorder realizations,
	we consider again the relative gain factor introduced
  in Section~\ref{subsubsec:numres-relative-gain} and defined in Eqs.~\eqref{eq:def-gain-med}
  and~\eqref{eq:def-gain-min}.
  The results are summarized in Table~\ref{tab:Glass_Instance_Gain}, 
  while a more general overview is provided in Figure~\ref{fig:EA-relgain},
  showing the distribution of relative gain factors for the 10 disorder realizations considered.
	As observed for the TFI model in Table~\ref{tab:gain_L3}, 
  regarding best runs ($\mathcal{G}_{\min}$) hard-ortho representation provides 
  no tangible advantage with respect to standard VQE, 
  even underperforming on several disorder realizations.
	Soft-ortho, instead, shows a clear improvement for all disorder realizations,
	with median gain factors fluctuating around 2.5
  for both versions ($\mathcal{G}_{\rm med}$ and $\mathcal{G}_{\min}$).
	For reference, the realization shown in Figure~\ref{fig:fiditglass}
  (labeled as `10' in Tables~\ref{tab:Glass_Instance_Gain})
  is on the mid-lower side of the gains distribution of Figure~\ref{fig:EA-relgain},
	with $\mathcal{G}_{\rm med} = 1.7(2)$ and $\mathcal{G}_{\min} = 2.30$.
  The relative gain factors achieved by the soft-ortho representation 
  on the EA model on a square $4\times 4$ lattice are lower than those
  observed for the TFI model in Table~\ref{tab:gain_L3} on a $3\times 3$ lattice,
  likely due to the increased complexity of the energy landscape of the former model,
  which makes the optimization more challenging overall.
  However, the results shown here confirm the effectiveness of the soft-ortho
  subspace representation also for larger and more challenging disordered models,
  where both standard VQE and hard-ortho approaches struggle to provide 
  accurate ground state estimates.

\section{Discussion and Conclusions}\label{sec:concl}

For both models numerically investigated in this work,
  we identified a non-monotonic behavior of fidelity metrics during the optimization
  (see Figures~\ref{fig:fiditising} and~\ref{fig:fiditglass}),
  signaling a possible underfitting phenomenon where the variational ansatz
  is not expressive enough to represent accurately the ground state
  but the cost function minimization steers the parameters
  towards regions of the parameter space where the overlap with 
  eigenstates just above the ground state is favored 
  (see Figure~\ref{fig:spectrograms-ising}).
  This issue affects both standard VQE and hard-ortho subspace representations,
  while the soft-ortho approach seems to mitigate it,
  providing a more consistent convergence behavior and final accuracies.
  We connect these improvements to the increased expressibility
  granted by the soft-ortho representation, 
  as argued in Section~\ref{subsec:subspace-express}.
  For the same reason, the typical circuit depths needed to reach final accuracies 
  comparable or even higher than standard VQE 
  are greatly reduced when using soft-ortho representations,
  as evident from Figures~\ref{fig:fiditising_boxplot} and~\ref{fig:fidvsparams_ising}.
  In the NISQ era, this is a relevant advantage,
  since it allows mitigating the detrimental effects of decoherence
  by employing relatively shallow circuits to represent the subspace frame states.
  Merely extending the search space from 
  subspace dimension $K=1$ (standard VQE) to $K=2$ significantly improves
  the accuracy of the ground state estimate at the end of the optimization
  and after diagonalization into the optimized subspace. 
  However, we also show that the hard-ortho representation, 
  employed in the SSVQE~\cite{PhysRevResearch.1.033062} and MCVQE~\cite{PhysRevLett.122.230401} approaches, 
  does not provide significant improvements over standard VQE 
  in reaching sufficiently high fidelities for the ground state estimates,
  while the soft-ortho representation systematically outperforms them.
  On the other hand, increasing the subspace dimension 
  beyond $K=2$ does not seem to improve the fidelity 
  and should be traded off with the total optimization time 
  and the risk of convergence issues within a fixed number of iterations.

As a future direction, it would be interesting to analyze
quantitatively the expressibility of subspace representations, 
possibly starting from well-known metrics for single-state ans\"atze 
from literature~\cite{ExpressibilitySim2019,Funcke2021dimensional,Arrasmith_2022,PRXQuantum.3.010313,Brozzi:2025omu}
and extending them to the subspace case.
Another possibly relevant aspect to investigate would be the effect 
of sampling noise and quantum noise, 
affecting independently both the evaluation of the cost function
during the optimization and the final diagonalization stage,
and impacting the accuracies of final ground state estimates 
and the relative gains (defined in Section~\ref{subsubsec:numres-relative-gain})
with respect to standard VQE.
Finally, it is possible to consider variations of 
the algorithm employing the soft-ortho representation 
as described in Section~\ref{subsubsec:subspace-soft}, 
exploring adaptive schemes with dynamic adjustments 
of the subspace dimension $K$ (i.e., adding/removing states)
or varying the orthogonality penalty parameter $\beta$ during the optimization.
It is also worth exploring the combination of well-established 
ansatz adaptive design techniques (such as ADAPT-VQE~\cite{grimsley2019adaptive})
with the soft-coded orthogonal subspace representation introduced in this work.

\acknowledgments
GC acknowledges support from the National Centre on HPC, Big Data and Quantum Computing -
SPOKE 10 (Quantum Computing) and received funding from the European Union Next-GenerationEU -
National Recovery and Resilience Plan (NRRP) – MISSION 4 COMPONENT 2, INVESTMENT N. 1.4 –
CUP N. I53C22000690001.
Numerical simulations have been performed on the \texttt{Leonardo} machines at CINECA, based on
the agreement between INFN and CINECA, under project INF25\_npqcd.

\bibliographystyle{apsrev4-1}
\bibliography{refs}

\end{document}